\documentclass[12pt]{article}

\usepackage{graphicx}
\usepackage{amssymb}

\RequirePackage{amsthm,amsmath}
\RequirePackage[colorlinks,citecolor=blue,urlcolor=blue]{hyperref}
\RequirePackage{hypernat}

\usepackage[default]{jasa_harvard}    

\RequirePackage{amssymb}
\RequirePackage{amsmath}
\RequirePackage{graphicx, rotating}
\RequirePackage{multirow}

\usepackage[left=1.25in,right=1.25in,top=1.22in,bottom=1.0in]{geometry}
\usepackage{setspace}

\numberwithin{equation}{section}
\theoremstyle{plain}

\newtheorem{theorem}{Theorem}[section]
\newtheorem{lemma}{Lemma}[section]
\newtheorem{proposition}{Proposition}[section]

\theoremstyle{definition}
\newtheorem{definition}{Definition}[section]

\newcommand{\bsb}{\boldsymbol}
\newcommand{\bsbX}{{\boldsymbol{X}}}
\newcommand{\bsbx}{{\boldsymbol{x}}}
\newcommand{\bsby}{{\boldsymbol{y}}}
\newcommand{\bsbb}{{\boldsymbol{\beta}}}
\newcommand{\bsbg}{{\boldsymbol{\gamma}}}

\newcommand{\bsbH}{{\boldsymbol{H}}}
\newcommand{\bsbI}{{\boldsymbol{I}}}

\newcommand{\bsbSig}{{\boldsymbol{\Sigma}}}
\newcommand{\bsbxi}{{\boldsymbol{\xi}}}
\newcommand{\bsbD}{{\boldsymbol{D}}}

\newcommand{\bsbU}{{\boldsymbol{U}}}

\newcommand{\bsbA}{{\boldsymbol{A}}}

\newcommand{\bsbeps}{{\boldsymbol{\epsilon}}}

\newcommand{\bsbr}{{\boldsymbol{r}}}

\newcommand{\real}{\mathbb{R}}
\newcommand{\tran}{{\mathsf{T}}}
\newcommand{\dnorm}{{\mathcal{N}}}

\newcommand{\rd}{\,\mathrm{d}}

\usepackage{algorithm,algorithmic}
\usepackage{paralist}
\newcommand{\deliver}{{\bf deliver}}
\newcommand{\given}{{\bf given}}

\newcommand{\adj}{{\,\mathrm{adj}}}
\newcommand{\ols}{{\mathrm{ols}}}
\newcommand{\soft}{{\mathrm{soft}}}

\begin{document}

\title{Outlier Detection Using Nonconvex Penalized Regression }
\author{Yiyuan She and Art B. Owen}
\date{}
\maketitle

\thispagestyle{empty}
\onehalfspacing
\begin{center}
\textbf{Author's Footnote:}
\end{center}
Yiyuan She is Assistant Professor, Department of Statistics, Florida State University.
Mailing address: Department of Statistics, Florida State University, Tallahassee, FL 32306-4330 (email: \texttt{yshe@stat.fsu.edu}). Art B. Owen is Professor, Department of Statistics, Stanford University. Mailing address: Department of Statistics, Stanford University, Stanford, CA 94305.
(email: \texttt{owen@stanford.edu}).
This work was partially supported by NSF grants DMS-0604939
and DMS-0906056.
The authors are grateful to Len Stefanski and the anonymous reviewers for
helpful remarks and useful pointers to the literature.
The authors thank J.\ Wisnowski and Professor J.\ Simpson for sharing the S-PLUS code in~\citeasnoun{jjcode}.

\setcounter{page}{1}
\doublespacing
\begin{abstract}
This paper studies the outlier detection problem from the point of view of penalized regressions. The regression model adds one mean shift parameter for each of the $n$ data points. We then apply a regularization favoring a sparse vector of mean shift parameters.  The usual $L_1$ penalty yields a convex criterion, but we find that it fails to deliver a robust estimator.  The $L_1$ penalty corresponds to soft thresholding. We introduce a thresholding (denoted by $\Theta$) based iterative procedure for outlier detection ($\Theta$-IPOD). A version based on hard thresholding correctly identifies outliers on some hard test problems.  We describe the connection between $\Theta$-IPOD and $M$-estimators. Our proposed method has one tuning parameter with which to both identify outliers and estimate regression coefficients. A data-dependent choice can be made based on BIC. The tuned $\Theta$-IPOD shows outstanding performance in identifying  outliers in various situations in comparison to other existing approaches. In addition, we find that $\Theta$-IPOD is much faster than iteratively reweighted least squares for large data because each iteration costs at most $O(np)$ (and sometimes much less) avoiding an $O(np^2)$ least squares estimate. This methodology extends to high-dimensional modeling with $p\gg n$, if both the coefficient vector and the outlier pattern are sparse.
\end{abstract}


%
%

\section{Introduction}
\label{sec:bg}
Outliers are a pervasive problem in statistical data analysis.
Nonrigorously, outliers refer  to one or more observations that are different from the bulk of the data.
\citeasnoun**{Hampelbook} estimate that  a routine data set may contain about {1-10\%} (or more) outliers. Unfortunately, outliers often go unnoticed~\cite{Roussbook}, although they may have serious effects in estimation, inference, and  model selection~\cite{weisb}.
Perhaps the most popular statistical modeling method is ordinary least squares (OLS)
regression.  OLS
is very sensitive to outliers --- a single unusual observation may break  it down completely.
Our goal in this work is  outlier identification for regression models,
together with robust coefficient estimation.

We consider the linear regression model given by $\bsby = \bsbX\bsbb+\bsbeps$
where $\bsbX\in\real^{n\times p}$ is a fixed matrix of predictors,
$\bsbb\in\real^p$ is a fixed (unknown) coefficient vector and $\bsbeps\in\real^n$
is a random error vector.
The $i$'th case is written $y_i = \bsbx_i^\tran\bsbb+\epsilon_i$.

Suspected outliers are most commonly found by looking
at residuals $r_i = y_i - \bsbx_i^\tran \hat\bsbb$
where $\hat\bsbb$ is the OLS estimate of $\bsbb$.
It is well-known that such raw residuals can fail to detect outliers
at leverage points.
A better way to detect an outlier is the \emph{leave-one-out} approach~\cite{weisb}.
If the $i$'th case is suspected to be an outlier, then we
compute the externally studentized residual
\begin{equation}\label{testoneout}
t_i =\frac{y_i - \bsbx_i^\tran\hat\bsbb_{(i)}}{\hat\sigma_{(i)}(1+\bsbx_i^\tran(\bsbX_{(i)}^\tran\bsbX_{(i)})^{-1}\bsbx_i)^{1/2}}
\end{equation}
where $\bsbX_{(i)}$, $\hat\bsbb_{(i)}$ and $\hat\sigma_{(i)}$ are the
predictor matrix, coefficient estimate
and scale estimate respectively, based on $n-1$ observations, leaving out the $i$'th.
Large values $|t_i|>\eta$ are then taken to suggest that
observation $i$ is an outlier. The threshold $\eta=2.5$ \cite{Roussbook} is a reasonable choice.
If $\bsbeps\sim \dnorm(0,\sigma^2I)$ then $t_i\sim t_{(n-p-1)}$
and we can even attach a significance level to $|t_i|$.
After removing an apparent outlier
from the data, one then looks for others.

Studentized residuals, and other leave-one-out methods
such as  Cook's distance and DFFITS, are simple and effective when there is only one outlier.
When there are multiple outliers, these simple methods can fail.
Two phenomena have been remarked on.  In \emph{masking}, when
an outlying $i$'th case has been left out, the remaining outliers cause either a large
value of $\hat\sigma_{(i)}$ or a small value of
$|y_i-\bsbx_i^\tran\hat\bsbb_{(i)}|$, or both, and as a result observation
$i$ does not look like an outlier. Therefore, multiple outliers may mask each other
and  go undetected.
In \emph{swamping}, the effect of
outliers is to make
$|y_i-\bsbx_i^\tran\hat\bsbb_{(i)}|$ large for a non-outlying case $i$.
Swamping could lead one to delete good observations and becomes more serious in the presence of multiple outliers.



In this paper we take the studentized residual as
our starting point. The $t$-test for whether observation $i'$
is an outlier is the same as testing whether the parameter $\gamma$
is zero in the regression
$\bsby = \bsbX\bsbb + \gamma 1_{i=i'} +\bsbeps$.
Because we don't know which observations might be outliers,
we use a model
\begin{equation}\label{msomodel}
\bsby = \bsbX\bsbb + \bsbg +\bsbeps,\qquad \bsbeps\sim \dnorm(0, \sigma^2 \bsbI),
\end{equation}
in which the parameter $\gamma_i$ is nonzero when observation $i$
is an outlier.
This formulation was earlier used by~\citeasnoun{gannaz} and~\citeasnoun{mcan}.
This mean-shift model allows any
combination of observations to be outliers.
It has $n+p$ regression  parameters and only $n$ data points.
Our approach is to fit~\eqref{msomodel} imposing
sparsity on $\bsbg$ in order to avoid the trivial
estimate $\hat\bsbg=\bsby$ and to get a meaningful
estimate of $\bsbb$.
The resulting algorithm is called thresholding (denoted by $\Theta$) based iterative procedure for
outlier detection, or $\Theta$-IPOD for short.

All of our proposals (apart from one exception noted where
it arises) require a preliminary robust regression to be run.
This practice is in line with the best current robust regression  methods.
The preliminary regression supplies a robust estimate
of $\bsbb$, and usually a robust estimate of $\sigma$ as well.
The robust regression methods that we compare to are known
to outperform the preliminary regressions that they use as
starting points. We will compare $\Theta$-IPOD to those best
performing methods.

The rest of the paper is organized as follows.
Section~\ref{sec:survey} surveys the literature
on robust regression.
Section~\ref{sec:softipod} develops the soft-IPOD
algorithm which fits~\eqref{msomodel} using an $L_1$
penalty on $\bsbg$.
This algorithm minimizes
a convex criterion, but it is \emph{not} robust.
Section~\ref{sec:thetaipod} develops a
family of  algorithms replacing soft-thresholding
by a general thresholding rule $\Theta$.
We find that some nonconvex criteria properly identify multiple
outliers in some standard challenging test cases.
Section~\ref{sec:compu} investigates the computational efficiency of $\Theta$-IPOD in comparison to iteratively reweighted least squares (IRLS).
IRLS requires a QR decomposition at each iteration,
while $\Theta$-IPOD needs only one.  As a result, though possibly requiring more iterations, it is much
faster in large problems that we investigate.
In Section~\ref{sec:tuning}, we discuss the important problem of parameter tuning in outlier detection and carry out an empirical study to demonstrate the advantage of our penalized approach.
Section~\ref{sec:largep} extends the technique to high-dimensional data
with $p>n$.
Our conclusions are in Section~\ref{sec:discussion}.

\section{Survey of robust regression}\label{sec:survey}
Many methods have been developed for multiple outlier identification.
Robust or resistant regressions, such as $M$-estimators~\cite{Huberbook}  and Least Trimmed Squares (LTS) \cite{Roussbook}, were proposed to provide a trustworthy coefficient estimate even in the presence of multiple outliers. There are also clustering-based procedures, such as~\citeasnoun**{serbert}, and some informal methods based on graphics. 
Multiple outlier detection procedures usually alternate between \textit{two}
steps.
One step scans regression output (coefficient and variance estimates)
to identify seemingly clean observations. The other fits a linear regression
model to those clean observations. The algorithm can be initialized with
OLS, but generally it is better to initialize it with something more robust.

If a data set contains more than one outlier, masking may occur and the task of outlier detection is much more challenging.
Well-known examples of masking include  
the Belgian telephone data  and the Hertzsprung-Russell star data \cite{Roussbook}, as well as some artificial datasets like the Hawkins-Bradu-Kass (HBK) data~\cite{HBK} and the Hadi-Simonoff (HS) data~\cite{HS}.
%
The HS and HBK data sets both have swamping effects.

The main challenge in multiple outlier detection is to counter masking and swamping effects.
\citeasnoun{HS} describe two broad classes of algorithms
--- direct methods and  indirect methods.
The direct procedures include  forward search algorithms~\cite{HS97,PY95,atki:rian:2000} and backward selection~\cite{menj:wels:2010} among others.
Indirect methods are those that use residuals from a robust
regression estimate to identify the outliers.
Examples of indirect methods include
Least Median of Squares (LMS)~\cite{lms}, Least Trimmed Squares (LTS)~\cite{Roussbook}, S-estimators~\cite{sest}, MM-estimators~\cite{mmest}, one-step GM estimators~\cite**{gmest} and S1S estimators~\cite{s1s}.
Almost all start with an initial high breakdown point  estimate (not necessarily efficient) say from LTS, S, MTS~\cite{Nguyen2010}, or~\possessivecite{PYfast} (denoted by PY) fast procedure.
Most published examples in the outlier
detection literature are in $10$ or fewer dimensions.
Methods with high breakdown point typically have costs that grow
exponentially in the dimension.
In practice, when there are a large number of predictors, PY can be applied to provide an initial estimate with certain robustness, although in theory its breakdown point property is not  well established~\cite{PYfast}.

It is worth mentioning that outlier identification and robust regression are two closely related but not quite identical problems~\cite**{Roussbook,Yohaibook}. 
Even if we perfectly identified the regression coefficient vector, there could still be some overlap between the residual distributions for good and outlying points. That is, given the true regression coefficient $\bsbb$ there would be type one and type two errors in trying to identify outliers. On the other hand, if we could perfectly identify all gross outliers, it is a relatively easy task to obtain a robust coefficient estimate, as will be supported by our experiments in  Section \ref{sec:tuning}.


%

\section{Soft-IPOD}\label{sec:softipod}
We will use the mean shift model~\eqref{msomodel} from the introduction
which predicts $\bsby$ by  the usual linear model $\bsbX\bsbb$
plus an outlier term $\bsbg$.
If $\gamma_i=0$ then the $i$'th case is good, and otherwise
it is an outlier. Our goals are to find a robust
estimate of $\bsbb$ as well as to estimate $\bsbg$
thereby identifying which cases are outliers and which are not.
We assume that $\bsbg$ is sparse because outliers should
not be the norm.
We suppose at first that $n>p$ and that $\bsbX = [\bsbx_1,\dots,\bsbx_n]^\tran$
has full rank $p$.
Section~\ref{sec:largep} considers the case when $\bsbb$ is sparse, too, with $p$ possibly greater than $n$. Yet the majority of our paper focuses on the outlier problem only. Let $\bsbH$ be the hat matrix defined by $\bsbH=\bsbH(\bsbX)=\bsbX(\bsbX^\tran\bsbX)^{-1}\bsbX^\tran$.
The $i$'th diagonal entry of $\bsbH$, denoted $h_i$, is called the leverage  of the $i$'th observation.

The assumed sparsity of $\bsbg$ motivates using an $L_1$-penalized regression
to minimize
\begin{eqnarray}\label{softopt}
f_\soft(\bsbb,\bsbg;\bsb{\lambda})\equiv
 \frac{1}{2} \| \bsby - \bsbX \bsbb - \bsbg\|_2^2 + \sum_{i=1}^{n} \lambda_i | \gamma_i|
\end{eqnarray}
over $\bsbb$ and $\bsbg$. The choice of $\bsb{\lambda}\in[0,\infty)^n$ is discussed below.
The function $f_\soft$ in \eqref{softopt} is jointly convex in $\bsbb$ and $\bsbg$.
Its simple form suggests that an alternating optimization can be applied:
given $\bsbg$, the optimal $\bsbb$ is the OLS estimate from regression of $\bsby-\bsbg$ on $\bsbX$;
given $\bsbb$, this $L_1$-penalized problem  is orthogonal and separable in $\bsbg$, and the optimal $\bsbg$ can be obtained by soft-thresholding.
This alternation between updates to $\bsbb$ and $\bsbg$ closely matches the interplay
between detecting outliers and fitting regression parameters mentioned in Section~\ref{sec:survey}.
We call the resulting algorithm soft-IPOD because it is based on soft thresholding.

We can derive a principled choice of $\bsb{\lambda}$ by comparing to
\possessivecite{Donoho} classical work which states that $\lambda(n) = \sigma \sqrt{2 \log n}$ is minimax optimal
when $\sigma$ is known and the predictors are orthogonal so that the residuals are uncorrelated.
Our residuals are correlated. For example, at the first step
 $\bsbr = \bsby - \bsbX\hat\bsbb_{\ols}\sim \dnorm(\bsbH\bsbg, \sigma^2 (\bsbI-\bsbH))$.
This motivates taking $\lambda_i=\sigma\sqrt{2(1-h_i)\log n}$ and then
{{\v{S}id\'{a}k's }inequality~\cite{Sid} controls the probability of wrongly
declaring $\gamma_i\ne 0$. It is conservative.
If there are many outliers, then $\sigma$ is not easily found.
As a result, we prefer to set $\lambda_i=\lambda\sqrt{1-h_i}$ and tune the regularization parameter $\lambda$ in a data-dependent way. See  Sections~\ref{sec:thetaipod} and~\ref{sec:tuning} for details.

To initialize the algorithm we must specify either $\bsbb$ or $\bsbg$.
Because $f_\soft$ is convex in $(\bsbb,\bsbg)$ the starting point is not crucial.
We could use $\bsbg=\bsb{0}$ for example. (For a detailed discussion of the initial estimate in the general situation, see Section \ref{sec:tuning} and Section \ref{sec:discussion}.)
A pseudo-code outline for soft-IPOD
is deferred to Section~\ref{sec:thetaipod}.
Soft-IPOD corresponds to Algorithm~\ref{alg:thetaipod} ($\Theta$-IPOD)
there, with a soft thresholding rule $\Theta$.
Soft-IPOD is guaranteed to converge, even though it is not contractive.
See Section \ref{subsec:ipodtisp}. 
There are other alternatives in computation for solving the convex optimization problem or its variants, such as the LARS~\cite**{Efron} used in \citeasnoun{mcan}. However, we shall see the Soft-IPOD offers a great advantage in generalizing the methodology to nonconvex penalized least-squares.

Although \eqref{softopt} is a well-formulated model and soft-IPOD is computationally efficient, in the presence of multiple outliers with moderate or high leverage values, this method fails to remove masking and swamping effects.
Take the artificial HBK data as an illustration. Using a robust estimate $\hat \sigma$ from  LTS, and $\lambda_i=\hat\sigma\sqrt{2(1-h_i)\log n}$, we obtained $\hat\bsbg=[0, \cdots, 0, -8.6,   -9.7,   -7.6,   -8.4, 0, \cdots, 0]^\tran$ which identifies cases 11-14 as serious outliers, while the true $\bsbg$ is $[10, \cdots, 10, 0, \cdots, 0]^\tran$ with cases 1-10 being the actual outliers.
This erroneous identification is not a matter of parameter tuning. Figure \ref{figHBKSoftPath}  plots the soft-IPOD solution over the relevant range for $\lambda$. Whatever value of $\lambda$ we choose, cases 11-14 are sure to be swamped, and cases 1-10 are very likely to be masked. The $L_1$ approach is not able to identify the correct outliers without swamping.
Our extensive experience shows that this $L_1$ technique hardly works for any benchmark dataset in the outlier detection literature.
According to \citeasnoun{Roussbook},  a convex criterion is inherently incompatible with robustness.

\begin{figure}[h]
\centering
\includegraphics[width=1.0\hsize, height=10cm]{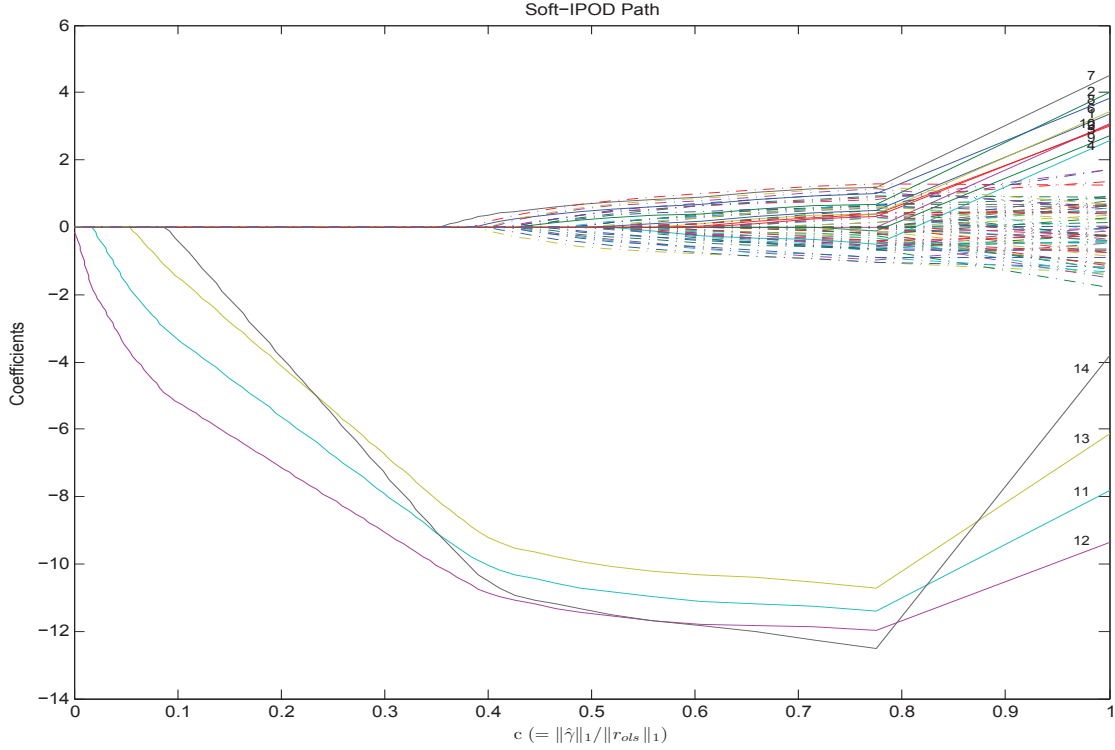}
\caption[]{\small{The solution path of the $L_1$ penalized regression on the HBK data. The estimates  $\hat\gamma_i(\lambda)$ ($i=1, 2, \cdots, 75$) are plotted against the constraint value $c$, given by $\|\hat\bsbg\|_1/\|\bsbr_{\mathrm{ols}}\|_1$ (the dual parameter of $\lambda$), where $\bsbr_{\mathrm{ols}}$ is the OLS residual vector. The top 10 paths correspond to cases 1-10, the true outliers. The bottom four correspond to cases 11-14 which are all good leverage points, but possibly swamped. (\textit{Figures are available in color online.})}}
\label{figHBKSoftPath}
\end{figure}

Next we delineate the parallels between soft-IPOD and Huber's M-estimate regression which is similarly non-robust. As pointed out by~\citeasnoun{gannaz} and~\citeasnoun{antrev}, there is a connection between the $L_1$-penalized regression \eqref{softopt} and Huber's $M$-estimate.
Huber's loss function is
$$
\rho(t;\lambda)=\begin{cases} \lambda |t| - \lambda^2/2, &\mbox{if } |t|>\lambda \\ t^2/2, &\mbox{if } |t|\le\lambda.  \end{cases}
$$
Huber's method with concomitant scale estimation, minimizes
\begin{align}\label{eq:huberslossconcomitant}
l(\bsbb, \sigma) \equiv\sum_{i=1}^n \rho\biggl(\frac{y_i - \bsbx_i^\tran\bsbb}\sigma;\lambda\biggr)\sigma
+ \frac{1}{2}  c n  \sigma
\end{align}
over $\bsbb$ and $\sigma$ jointly, where $c\ge0$ and $\lambda > 0$
are given constants.
We can prove the following result, which is slightly more general than
the version given by~\citeasnoun{gannaz} and~\citeasnoun{antrev}.
\begin{proposition}
\label{propl1huber}
For any $c\ge 0$ and $\lambda>0$ suppose that we minimize
\begin{eqnarray}\label{soft-ipodopt}
 g(\bsbb, \bsbg, \sigma) \equiv\frac{1}{2\sigma} \left\|  \bsby - \bsbX \bsbb-\bsbg \right\|_2^2+ \frac{1}{2}\, c n \sigma + \lambda \| \bsbg \|_1,
\end{eqnarray}
over $\bsbb$, $\bsbg$ and $\sigma$. Then the minimizing $\bsbb$ and $\sigma$
match those from minimizing Huber's criterion~\eqref{eq:huberslossconcomitant}.
For any fixed $\sigma>0$ minimizing~\eqref{soft-ipodopt} over $\bsbb$ and $\bsbg$
yields the minimizer of~\eqref{eq:huberslossconcomitant} over $\bsbb$.
\end{proposition}

This connection is helpful to understand the inherent difficulty with $L_1$-penalized regression described earlier. It is well known that Huber's method cannot even handle moderate leverage points well~\cite[p. 192]{Huberbook}
and is prone to masking and swamping in outlier detection. Its break-down point  is~$0$. 

A promising way to improve \eqref{softopt} is to adopt a different penalty function, possibly nonconvex. Our launching point in this paper is, however, the operator $\Theta$ in the above iterative algorithm. Substituting  an appropriate $\Theta$ for soft-thresholding in the IPOD (see Algorithm~\ref{alg:thetaipod}), we may obtain a good estimator of $\bsbg$ (and $\bsbb$ as well).

\section{$\Theta$-IPOD}
\label{sec:thetaipod}
To deal with masking and swamping in the presence of multiple outliers, we consider the iterative procedure described in Section \ref{sec:softipod} using a general $\Theta$ operator, referred to as
$\Theta$-IPOD. A $\Theta$-IPOD estimate is a limit point of $(\bsbb^{(j)},\bsbg^{(j)})$, denoted by $(\hat\bsbb, \hat\bsbg)$.
Somewhat surprisingly, simply replacing soft-thresholding by hard-thresholding   (henceforth hard-IPOD)
resolves the masking and swamping problem, for the challenging HBK example problem.
After picking $\hat\sigma$ from LTS and using a zero start
we obtained
\begin{align*}
\hat\bsbg_{1:10} &=\begin{pmatrix}9.7& 10.2&    10.4&     9.7&    10.1&     10.0&    10.8&   10.4&     9.8&    10.1\end{pmatrix}^\tran
\end{align*}
and $\hat\bsbg_{11:75}=\bsb{0}$
which perfectly detects the true outliers. The coefficient estimate $\hat\bsbb$ from hard-IPOD
directly gives the OLS estimate computed from the clean observations.

Some questions naturally arise:  What optimization problem is $\Theta$-IPOD trying to solve?  For an arbitrary $\Theta$, does $\Theta$-IPOD converge at all? Or, under what conditions does $\Theta$-IPOD converge? To answer these questions, we limit our discussions of $\Theta$ to thresholding rules.

\subsection{IPOD, Penalized Regressions, and $M$-estimators}
\label{subsec:ipodpenM}
We begin by defining the class of threshold functions we will study.
It includes well-known thresholds such as soft and hard thresholding,
SCAD, and Tukey's bisquare.

\begin{definition}[Threshold function]\label{def:threshold}
A threshold function is a real valued
function  $\Theta(t;\lambda)$ defined for $-\infty<t<\infty$
with $\lambda$ as the parameter ($0\le\lambda<\infty$) such that
\begin{compactenum}[\qquad\bf 1)]
\item $\Theta(-t;\lambda)= -\Theta(t;\lambda)$,
\item $\Theta(t;\lambda)\le \Theta(t';\lambda)$ for $t\le t'$, 
\item $\lim_{t\to\infty} \Theta(t;\lambda)=\infty$,\quad and
\item $0\le \Theta(t;\lambda)\le t$\ for\ $0\le t<\infty$.
\end{compactenum}
In words, $\Theta(\cdot;\lambda)$  is an odd monotone unbounded shrinkage
rule for $t$, at any $\lambda$.
\end{definition}
A vector version of $\Theta$ is defined componentwise if
either $t$ or $\lambda$ are replaced by vectors. When both
$t$ and $\lambda$ are vectors, we assume they have the same
dimension.

Huber's soft-thresholding rule corresponds to an absolute error, or $L_1$  penalty.
More generally, for any thresholding rule $\Theta(\cdot; \lambda)$,
a corresponding penalty function $P=P_\Theta$ can be defined.
There may be multiple penalty functions for a given threshold
as demonstrated by 
hard thresholding \eqref{eq:hardthresh} below.
The following \emph{three-step construction} finds the penalty with the smallest curvature~\cite{antrev,SheTISP}:
\begin{equation}\label{defofthetainv}
\begin{split}
\Theta^{-1}(u;\lambda)&=\sup\{t:\Theta(t;\lambda)\leq u\},  
\\
s(u;\lambda) &=  \Theta^{-1}(u;\lambda)-u,\quad\text{and} \\
P(\theta;\lambda)&=P_\Theta(\theta;\lambda)=\int_0^{|\theta|} s(u;\lambda)\rd u,
\end{split}
\end{equation}
where $u\ge 0$ holds throughout~\eqref{defofthetainv}.
The constructed penalty $P_\Theta$ is nonnegative and is continuous in $\theta$.

\begin{algorithm}[t]
{
\begin{algorithmic}
\caption{$\Theta$-IPOD\strut\label{alg:thetaipod}}
\STATE \given\ $\bsbX\in\real^{n\times p}$,\  $\bsby\in\real^n$,\  $\varepsilon>0$,\  $\bsb{\lambda}\in[0,\infty)^n$,\
robust pilot estimate $\bsbb^{(0)}\in\real^p$
\STATE $j\gets 0$, $\bsbg^{(j)}\gets\bsby-\bsbX\hat\bsbb$, $\mathrm{converged}\gets\mathrm{FALSE}$
\STATE $(Q,R)\gets\mathrm{QRdecomp}(\bsbX)$
\STATE \qquad// $\bsbX=QR$, $R\in\real^{p\times p}$ is upper triangular, $Q\in\real^{n\times p}$ and $Q^\tran Q=I_p$
\WHILE{not converged}
\STATE $\bsby^\adj \gets\bsby-\bsbg^{(j)}$
\STATE $\bsbb^{(j)} \gets  R^{-1}Q^\tran\bsby^\adj$
\STATE $\bsbr^{(j)} \gets\bsby - \bsbX \bsbb^{(j)}$
\STATE $\bsbg^{(j+1)}\gets{\Theta}(\bsbr^{(j)}; \bsb{\lambda})$
\STATE $\mathrm{converged}\gets \Vert\bsbg^{(j+1)}-\bsbg^{(j)}\Vert_\infty<\varepsilon$
\STATE $j\gets j+1$
\ENDWHILE 
\STATE\deliver\ $\hat\bsbb=\bsbb^{(j-1)}$,\quad $\hat\bsbg = \bsbg^{(j)}$
\end{algorithmic}
\vspace*{0.1cm}
}
\end{algorithm}

\begin{theorem}\label{thm:itsdecreasing}
\label{thpen}
Let $\Theta$ be a thresholding rule as given by Definition~\ref{def:threshold}
and let $P_\Theta$ be the corresponding penalty defined in~\eqref{defofthetainv}.
Given $\lambda_i\geq 0$, the objective function is defined by 
\begin{eqnarray}
f_P(\bsbb, \bsbg)
\equiv \frac{1}{2} \|\bsby-\bsbX \bsbb- \bsbg\|_2^2 + \sum_{i=1}^n P(\gamma_i; \lambda_i), \label{ipodopt}
\end{eqnarray}
where  $P(\cdot;\cdot)$  is any function satisfying
$P(\theta;\lambda)-P(0;\lambda)=P_\Theta(\theta; \lambda) + q(\theta; \lambda)$ where $q(\cdot;\lambda)$ is nonnegative and $q(\Theta(\theta;\lambda))=0$ for all $\theta$.
Then the $\Theta$-IPOD iteration sequence $(\bsbb^{(j)},\bsbg^{(j)})$ satisfies
\begin{eqnarray}
f_P(\bsbb^{(j)}, \bsbg^{(j)}) \geq f_P(\bsbb^{(j)}, \bsbg^{(j+1)}) \geq f_P(\bsbb^{(j+1)}, \bsbg^{(j+1)}),\quad \mbox{for any}\ j\ge 0. \label{ipodconvineq}
\end{eqnarray}
\end{theorem}

The function $q$ will
often be zero, but we use non-zero
$q$ 
below to demonstrate that multiple penalties (infinitely many, as a matter of fact)
yield hard thresholding, including the $L_0$-penalty.
Theorem~\ref{thm:itsdecreasing} shows that  $\Theta$-IPOD converges.
Any limit point of $(\bsbb^{(j)},\bsbg^{(j)})$ must be a stationary point of \eqref{ipodopt}.
$\Theta$-IPOD also gives a general connection between penalized regression~\eqref{ipodopt} and $M$-estimators. Recall that an $M$-estimator is defined to be a solution to the score equation
\begin{eqnarray}
\sum_{i=1}^n \psi \left( \frac{y_i - \bsbx_i^\tran \bsbb}\sigma; \lambda \right)\bsbx_{i} = \bsb{0}, \label{psieq}
\end{eqnarray}
where $\lambda$ is a general parameter of the $\psi$ function. Although $\bsbb$ and $\sigma$ can be simultaneously estimated by Huber's Proposal 2~\cite{Huberbook}, a more common practice is to fix $\sigma$ at an initial robust estimate and then optimize over $\bsbb$~\cite{Hampelbook}. Unless otherwise specified, we consider equation \eqref{psieq} as constraining $\bsbb$ with  $\sigma$ fixed.

\begin{proposition}
\label{ipodm}
For any thresholding rule $\Theta(\cdot; \lambda)$, for any $\Theta$-IPOD estimate $(\hat \bsbb, \hat\bsbg)$,  $\hat \bsbb$  is an $M$-estimate associated with $\psi$, as long as $(\Theta, \psi)$ satisfies
\begin{eqnarray}
\Theta(t; \lambda) + \psi(t; \lambda)=t,\quad \forall t. \label{iden}
\end{eqnarray}
\end{proposition}

We have mentioned that Huber's method or soft-IPOD
behaves poorly in outlier detection. The problem is that it never rejects gross outliers that have moderate or high leverage. To reject gross outliers, redescending $\psi$-functions are advocated, corresponding to a class of thresholdings offering little shrinkage for large components, or using nonconvex penalties for solving the sparsity problem \eqref{ipodopt}.
The differences between $\Theta$-IPOD and the corresponding $M$-estimator are as follows:
\begin{compactenum}[(i)]
\item $M$-estimators focus on robust estimation of $\bsbb$. For an explicit sparse $\bsbg$ estimate,  a cutoff value is usually needed for the residuals.  Minimizing $f_P$
from equation~\eqref{ipodopt} directly yields a sparse $\hat\bsbg$ for outlier detection and a robust~$\hat\bsbb$.
\item Instead of designing a robust loss function in $M$-estimators,  \eqref{ipodopt} considers a penalty function; $\lambda$ is not a criterion (loss) parameter but a regularization parameter that we will tune in a data-dependent way.
\end{compactenum}
Figure \ref{figfuncsdemon} illustrates some of the better
known threshold functions along with their corresponding
penalties, $\psi$-functions and loss functions.
In this article, we make use of the following formulas
\begin{align}
\Theta_{\mathrm{soft}}(x; \lambda)&=\begin{cases}0, & \mbox{if } |x|\leq \lambda\\ x - \mbox{sgn}(x)\lambda, & \mbox{if } |x|>\lambda,\end{cases}\label{eq:softthresh}\\
\Theta_{\mathrm{hard}}(x; \lambda)&=\begin{cases}0, &\mbox{if } |x|\leq \lambda\\ x, &\mbox{if } |x|>\lambda,\end{cases}\label{eq:hardthresh}
\end{align}
for soft and hard thresholding, respectively.
The usual penalty for hard thresholding
is $P(x;\lambda) =  1_{|x|< \lambda} (\lambda|x|-x^2/2)+  1_{|x|\ge \lambda} \lambda^2/2$. 
Theorem \ref{thpen} justifies use of
the $L_0$-penalty $P(x; \lambda)=\lambda^2/2 \cdot I_{x\neq 0}$
using
$$
q(x;\lambda)=\begin{cases} \frac{(\lambda-|x|)^2}{2} , & \mbox{ if } 0 < |x| < \lambda\\
0,  & \mbox{ if }  x=0 \mbox{ or } |x| \geq \lambda. \end{cases}
$$

\begin{figure}[h]
\begin{center}
\includegraphics[width=4.5in]{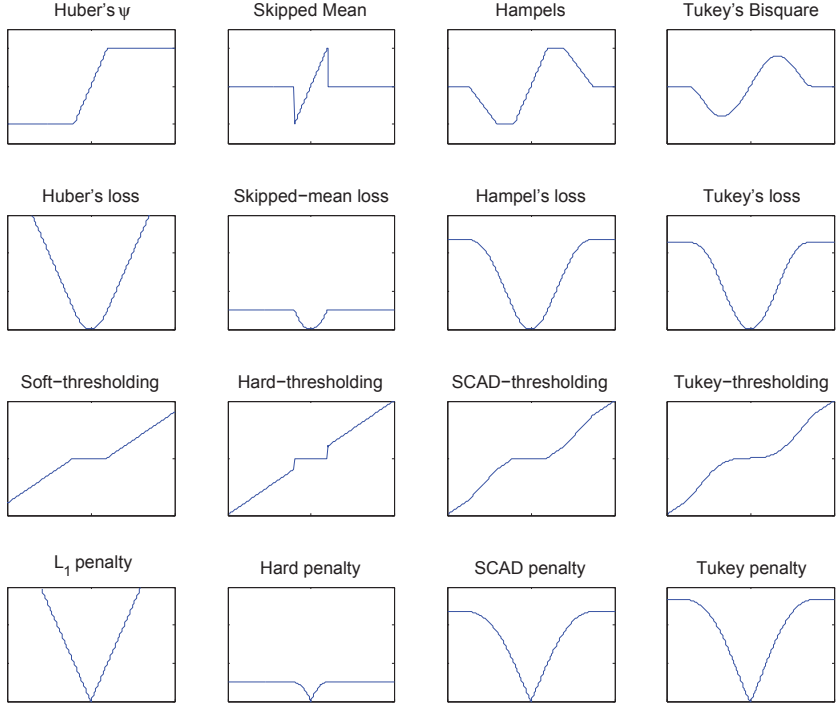}
\end{center}
\caption[]{\small{The $\psi$ functions (top row), loss functions (second row), thresholding functions (third row), and penalty functions (bottom row) for some well known thresholding rules.
SCAD-thresholding~\cite{AntFan} is a special case of
the Hampel rule.
}}
\label{figfuncsdemon}
\end{figure}


\subsection{$\Theta$-IPOD and TISP}
\label{subsec:ipodtisp}


Algorithm \ref{alg:thetaipod} can be simplified.
The $\Theta$-IPOD iteration can be carried out without recomputing $\bsbb^{(j)}$ at each iteration.
We need only update $\bsbg$ via
\begin{eqnarray}
\bsbg^{(j+1)} = \Theta(\bsbH \bsbg^{(j)} + (\bsbI-\bsbH)\bsby; \bsb{\lambda}), \label{simpIPOD}
\end{eqnarray}
at each iteration, where $\lambda_i=\lambda \sqrt{1-h_i}$.
The multiplication $(\bsbI-\bsbH)\bsby$ in~\eqref{simpIPOD} can also be precomputed.
After getting the final $\hat\bsbg$, we can estimate $\bsbb$ by OLS.
The resulting simplified $\Theta$-IPOD
algorithm is given in Algorithm~\ref{alg:thipod}.
For any given thresholding rule $\Theta$,  let
$f_P(\bsbg)\equiv \frac{1}{2} \|(\bsbI - \bsbH) (\bsby- \bsbg)\|_2^2 + \sum_{i=1}^n P(\gamma_i; \lambda_i),
$ where $P$ can be any penalty function satisfying the conditions in  Theorem \ref{thpen}. Then $f_P(\bsbg^{(j+1)})\leq f_P(\bsbg^{(j)})$ for the $\Theta$-IPOD iterates $\bsbg^{(j)}$, $j\ge 0$.

\begin{algorithm}[t]
\begin{algorithmic}
\caption{Simplified $\Theta$--IPOD\strut\label{alg:thipod}}
\STATE \given\  $\bsbX\in\real^{n\times p}$,   $\bsby\in\real^n$,  $\lambda>0$,  $\varepsilon>0$,   $\bsbg^{(0)}\in\real^p$,
and threshold $\Theta(\cdot;\cdot)$
\STATE $j\gets 0$, $\bsbg^{(j)} \gets \bsbg^{(0)}$, $\mathrm{converged}\gets\mathrm{FALSE}$
\STATE $\bsbH\gets\mathrm{HatMatrix}(\bsbX)$, $\bsbr\gets \bsby-\bsbH\bsby$
\WHILE{not converged}
\STATE $\bsbg^{(j+1)} \gets \Theta(\bsbH \bsbg^{(j)} + \bsbr; \bsb{\lambda})$
\STATE $\mathrm{converged}\gets \Vert\bsbg^{(j+1)}-\bsbg^{(j)}\Vert_\infty<\varepsilon$
\STATE $j\gets j+1$
\ENDWHILE
\STATE\deliver\ $\hat\bsbg = \bsbg^{(j)}$,\quad $\hat \bsbb = \mathrm{OLScoef}(\bsbX,\bsby-\hat\bsbg)$
\end{algorithmic}
\vspace*{.1cm}
$\bsbX$ must have full rank. $\bsbg^{(0)}$  can be
$\bsby-\bsbX\bsbb^{(0)}$
for a robust pilot estimate $\bsbb^{(0)}$.
\end{algorithm}

Setting up for Algorithm~\ref{alg:thipod} costs $O(np^2)$ for a dense
regression. The dominant cost in a given iteration comes from
computing $\bsbH\bsbg^{(j)}$. Given a QR decomposition of $\bsbH$
we can compute that matrix product in $O(np)$ work as $Q(Q^\tran \bsbg^{(j)})$.
The cost of the update could
be even less if $\bsbg^{(j)}$ has fewer than $p$ nonzero entries
and one maintains the dense matrix $\bsbH$.

To give \eqref{simpIPOD} another explanation, suppose the spectral decomposition of the hat matrix $\bsbH$ is given by $\bsbH=\bsbU \bsbD \bsbU^\tran$. Define an index set $c=\{i: D_{ii} = 0\}$ and let $\bsbU_c$ be formed by taking the corresponding columns of $\bsbU$.  Then a \emph{reduced model} can be obtained from the mean shift outlier model \eqref{msomodel}
\begin{eqnarray}\label{redmodel}
\tilde\bsby=\bsbA\bsbg +  {\bsbeps'},\quad \bsbeps'\sim \dnorm(\bsb{0}, \sigma^2\bsbI_{(n-p)\times (n-p)}),
\end{eqnarray}
where  $\tilde\bsby=\bsbU_c^\tran\bsby$ and $\bsbA = \bsbU_c^\tran \in \real^{(n-p)\times n}$.
The term $\bsbX\bsbb$ has disappeared from the reduced model. For a special $\bsbU_c$, the vector $\tilde\bsby$ has the BLUS residuals of~\citeasnoun{thei:1965}.
The regression model~\eqref{redmodel} is a $p>n$ sparsity problem.
It is also a wavelet approximation problem that~\citeasnoun{AntFan} studied in the context of wavelet denoising, because $\bsbA$ satisfies $\bsbA\bsbA^\tran= \bsbI$. Furthermore, the regularized one-step estimator (ROSE)~\cite{AntFan} is  the first step of $\Theta$-IPOD.

We can build a connection between the reduced model and simplified $\Theta$-IPOD. 
\citeasnoun{SheTISP} proposed a class of  thresholding-based iterative selection procedures (TISP) for model selection and shrinkage.
$\Theta$-TISP for solving the sparsity problem \eqref{redmodel} is given by
\begin{align}\label{eq:thetatispspar}
\bsbg^{(j+1)} = \Theta(\bsbA^\tran \tilde\bsby/k_0^2  + (\bsbI-\bsbA^\tran\bsbA/k_0^2)\bsbg^{(j)};
\bsb{\lambda}/k_0^2)
\end{align}
with  $k_0$ equal to the largest singular value of the Gram matrix $\bsbA^\tran\bsbA=\bsbH$.
The iteration~\eqref{eq:thetatispspar} reduces  exactly to \eqref{simpIPOD}.
Therefore, all TISP studies apply to the $\Theta$-IPOD algorithm.
For example, the TISP convergence theorem can be used to establish a version of Theorem \ref{thm:itsdecreasing}, and the nonasymptotic probability bounds for sparsity recovery reveal masking and swamping errors.
In particular, the advocated hard-thresholding-like $\Theta$ in TISP corresponds to a redescending $\psi$ in our outlier identification problem.

The simplified procedure \eqref{simpIPOD} is easy to implement and is computationally efficient, because the iteration does not involve complicated operations like matrix inversion. Model \eqref{redmodel} is simpler than the original \eqref{msomodel} because $\bsbb$ does not appear
and all observations are \emph{clean}. They have non-outlying errors
because we have moved the outlier variables into the regression.
Using this characterization of $\hat\bsbg$, it is not difficult to show that the IPOD-estimate $\hat\bsbb$  satisfies the regression, scale, and affine equivariant properties~\cite{Hampelbook} desirable for a good robust regression estimator:
\begin{compactenum}[\quad(i)]
\item $\hat\bsbb(\bsbX, \bsby + \bsbX\bsb{\eta})=\hat\bsbb(\bsbX, \bsby)+\bsb{\eta},\quad \forall \bsb{\eta}\in \real^p$;
\item $\hat\bsbb(\bsbX, c\bsby )=c\hat\bsbb(\bsbX, \bsby),\quad \forall c\in\real$;
\item $\hat\bsbb(\bsbX \bsb{C}, \bsby )=\bsb{C}^{-1} \hat\bsbb(\bsbX, \bsby)$,\quad for any nonsingular $\bsb{C}$.
\end{compactenum}
We are making here a mild assumption that the initial robust estimate is equivariant, as LTS, S and PY~\cite{PYfast} are, and we're ignoring the possible effect of convergence criteria on equivariance.

It remains to select $\lambda$.
The clean dataset $(\bsbA, \tilde \bsby)$ in \eqref{redmodel} has
uncorrelated noise contamination, 
 so we can use it to tune the parameters
via BIC and related methods; see Section \ref{sec:tuning} for details.

\section{$\Theta$-IPOD vs. IRLS}\label{sec:compu} 
We consider some computational issues in this section.
As Proposition \ref{ipodm} suggests, $\Theta$-IPOD solves an $M$-estimation problem. The standard fitting algorithm
for $M$-estimates is the well-known iteratively re-weighted least squares (IRLS).
Let $w(t;\lambda)=\psi(t;\lambda)/t$, taking $0/0=0$ if necessary.
The $\psi$-equation \eqref{psieq} that defines an $M$-estimator can be rewritten as
$\sum_{i=1}^n w\left({r_i}/\sigma; \lambda\right)  r_i \bsbx_{i} = \bsb{0}$,
where $r_i=y_i - \bsbx_i^\tran\bsbb$.  Accordingly, an $M$-estimate corresponds to a weighted LS estimate
as is well known.
These multiplicative weights can help downweight the bad observations.
Iteratively updating the weights yields the IRLS algorithm, which is the most common method for computing $M$-estimates.
Model \eqref{msomodel} indicates that $M$-estimation can also be characterized through \textbf{additive} effects on all observations.

We performed a simulation to compare the speed of $\Theta$-IPOD to that of IRLS.
The observations were generated according to $\bsby = \bsbX \bsbb + \bsbg + \bsbeps$, where $\bsbeps\sim \dnorm(0, \bsbI)$.
The predictor matrix $\bsbX$ is constructed as follows: first let $\bsbX= \bsbU \bsbSig^{1/2}$, where $U_{ij} \buildrel {\mathrm{iid}} \over  \sim~U(-15, 15)$ and $\bsbSig_{ij}=\rho^{1_{i\neq j}}$ with $\rho=0.5$; then modify the first $O$ rows to be leverage points given by $L\cdot [1, \cdots, 1]$.
We consider all nine cases  with $L\in\{15, 20, 30\}$ and
$O\in\{5,10,20\}$. Three more cases correspond to additive outliers at $O$ points that
were not leverage points (no rows of $\bsbX$ changed).
The shift vector is given by $\bsbg=(\{8\}^{O}, \{0\}^{n-O})^\tran$.

Given $n$ and $p$, we report the total cost of computing all  $M$-estimates for  these different combinations of  the number of outliers ($O$)
and leverage value ($L$), each combination simulated 10 times. The scale parameter ($\lambda$) decreased from $\|(\bsbI - \bsbH)\bsby./\sqrt{\mbox{diag}(\bsbI-\bsbH)}\|_\infty$ to $0.5$ with fixed step size $-0.1$,
where ./ stands for  elementwise division.  The upper limit is the largest possible standardized
residual. Empirically, the lower bound $0.5$  yields approximately half of $\gamma_i$ nonzero. Also for $\lambda<0.5$  IRLS often encountered a singular WLS  during the iteration, or took exceptionally long time to converge, and so could not be compared to $\Theta$-IPOD.
We used IRLS (with fixed $\sigma=1$, as an oracle would have)
and simplified $\Theta$-IPOD. The common convergence  criterion was $\Vert\bsbg^{(j+1)}-\bsbg^{(j)}\Vert_\infty<10^{-4}$.
We studied all sample sizes
 $n\in\{30, 50, 100, 200, 300, 400, 500, 600, 700, 800, 900, 1000\}$
and we took $p=n/10$.
The CPU times (in seconds)  are plotted against the sample size in Figure \ref{figcostcomp}.


\input{epsf}
\begin{figure}[t!]
\centering
$\begin{array}{c@{\hspace{0.0in}}c}
\epsfxsize=3.3in 
\epsffile{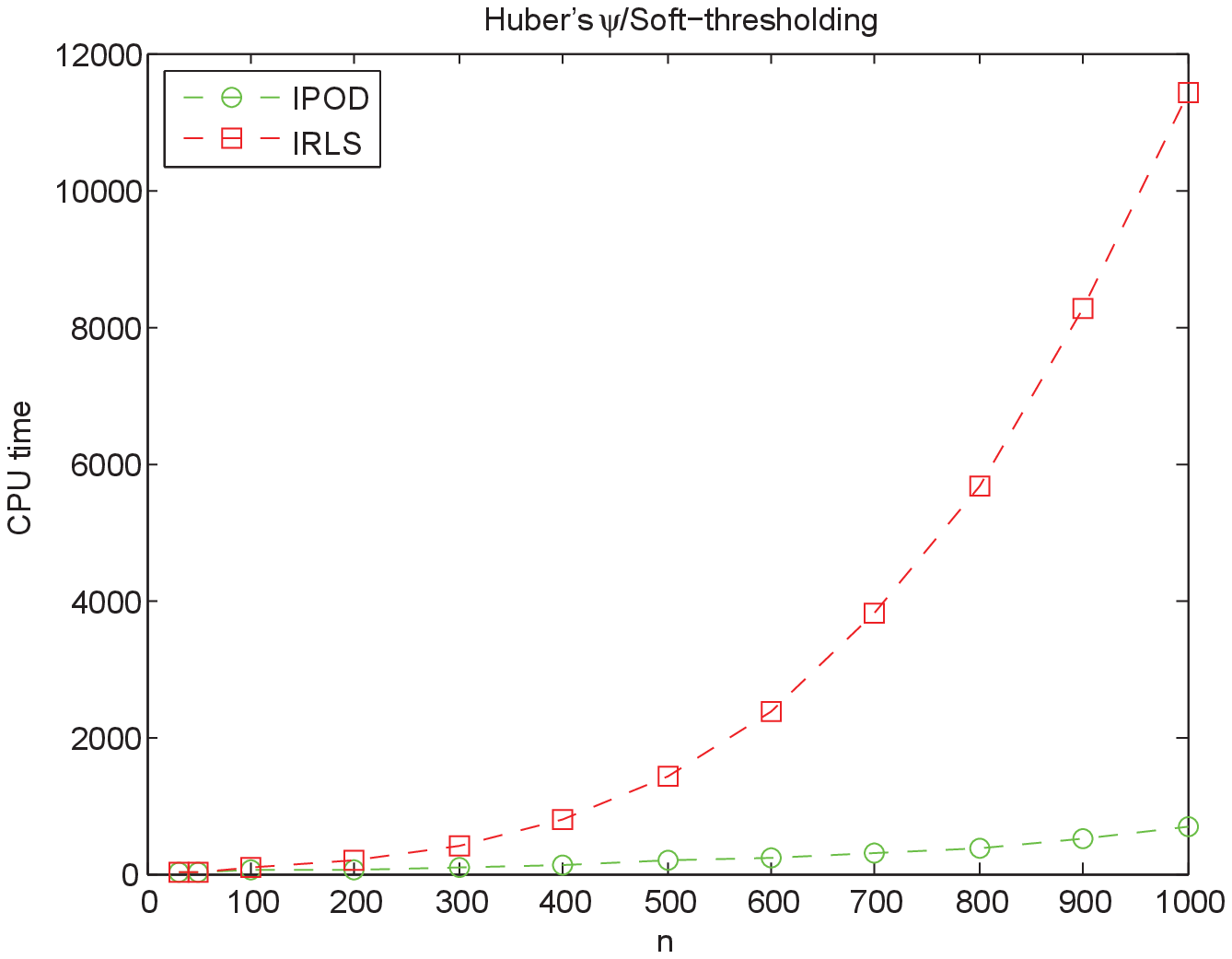} &
\epsfxsize=3.3in 
\epsffile{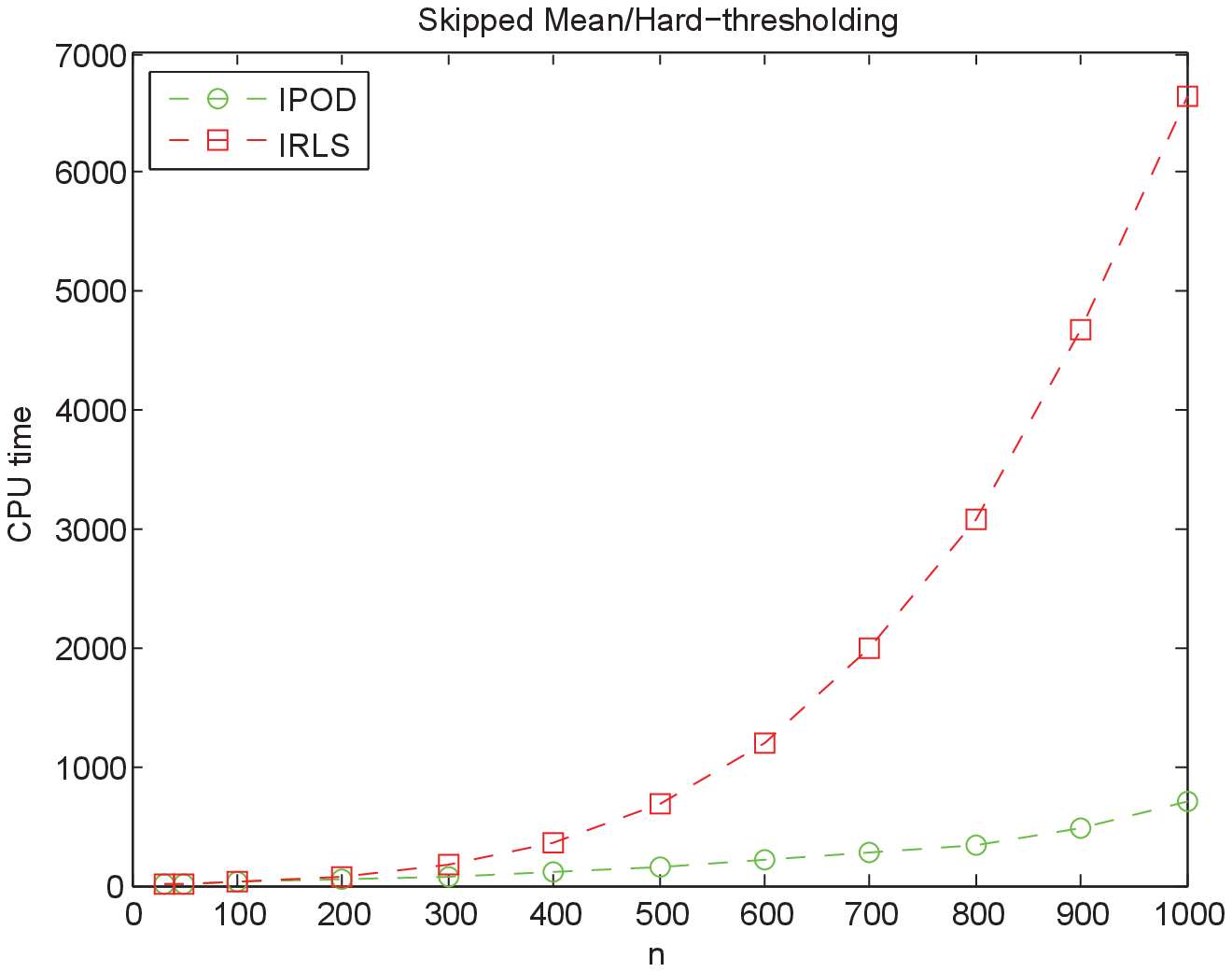} \\
\epsfxsize=3.3in 
\epsffile{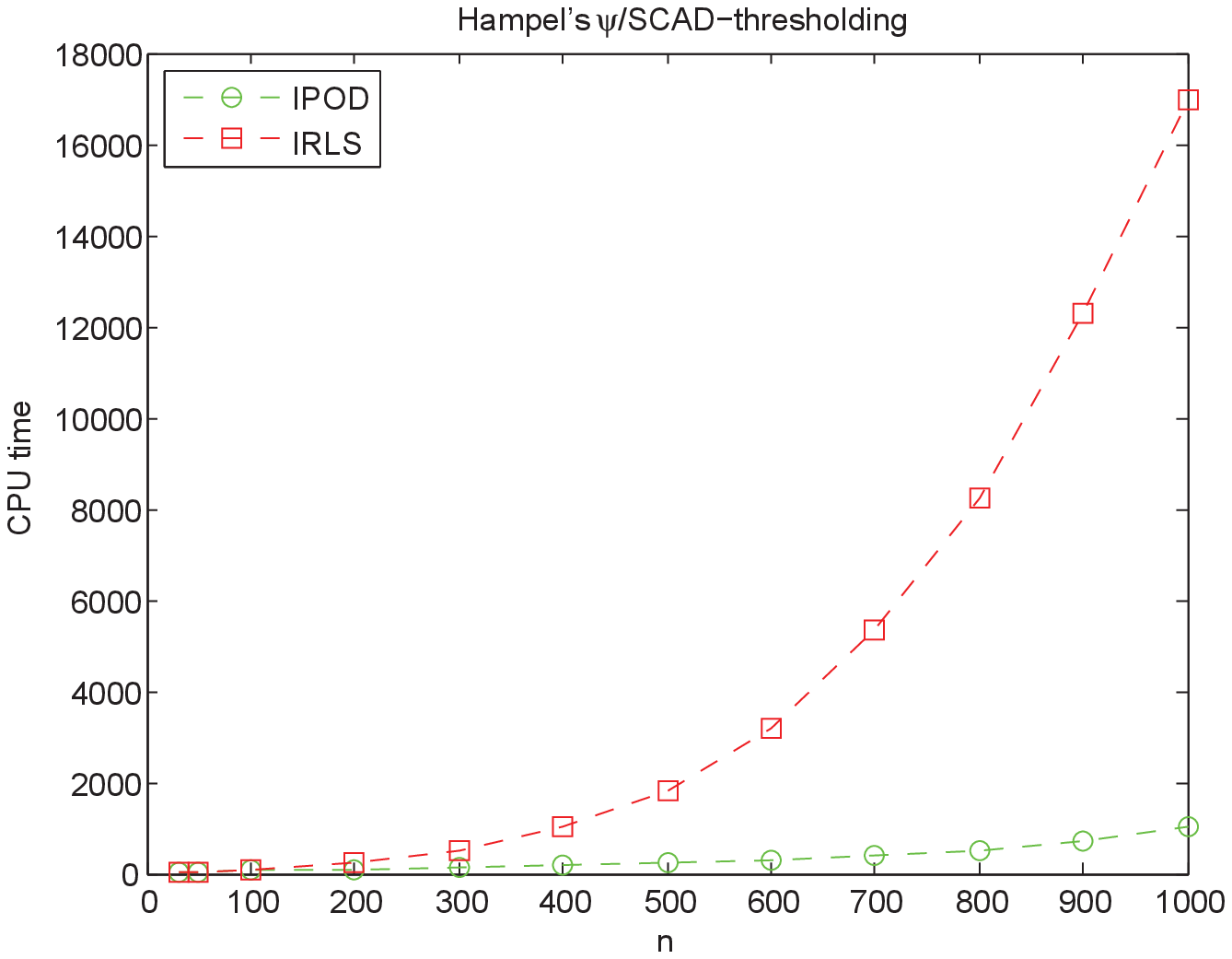} &
\epsfxsize=3.3in 
\epsffile{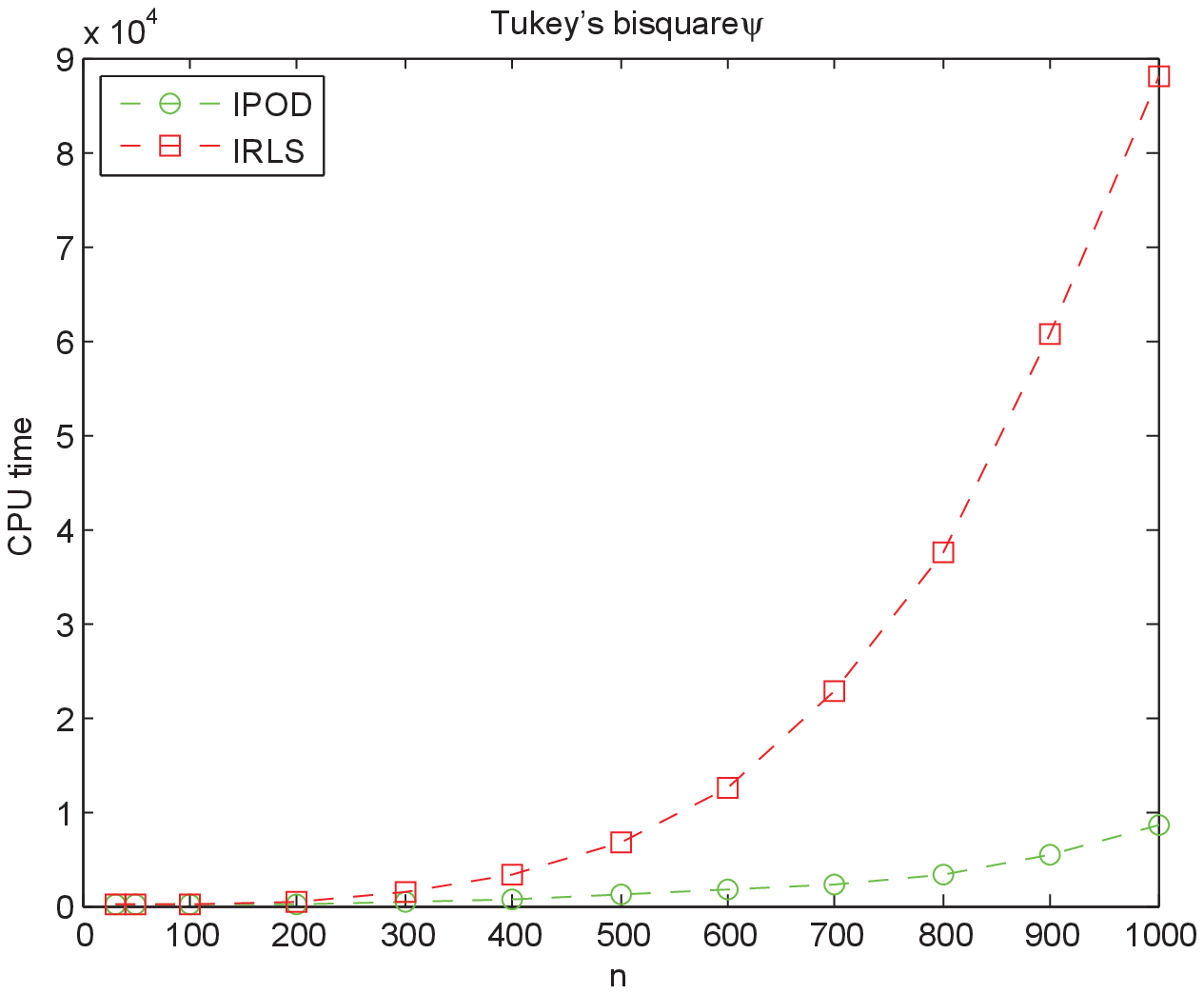}
\end{array}$
\caption[]{\small{Computational time comparison between $\Theta$-IPOD and IRLS for problems of different sizes.  The IRLS computational times are represented by red squares, and the $\Theta$-IPOD computational times by green circles. The plots, from left to right, top to bottom, correspond to the examples in Figure \ref{figfuncsdemon} 
respectively.}}\label{figcostcomp}
\end{figure}

The speed advantage of $\Theta$-IPOD over IRLS in
this simulation tended to increase with $n$ and $p$.
At $n=1000$ the hard-IPOD algorithm required about 2 or 3 times (averaged over all lambda values) as many iterations as IRLS but was about $10$ times faster than IRLS, due to faster iterations.
For small $n$, we saw little speed difference.
As remarked above, IRLS was sometimes unstable, with singular WLS problems arising for redescending $\psi$, when fewer than $p$ weights were nonzero.
Such cases cause only a small problem for $\Theta$-IPOD:
the update $\bsbH\bsbg^{(j)}$ cannot then take advantage of sparsity,
but it is very stable.

Although we attained impressive computational gain over the popular IRLS, we consider speed to be secondary compared to robustness (and the speed advantage of $\Theta$-IPOD can be moot when the preliminary method is very expensive).

\section{Parameter Tuning in Outlier Detection}\label{sec:tuning}
The parameter $\lambda$ in an $M$-estimators \eqref{psieq} is often chosen to be a constant (for all $n$),
based on either efficiency or breakdown of the estimator.
The value $2.5\hat\sigma$ is popular~\cite**{Roussbook,Wilbook,Yohaibook}.
But as mentioned in Section \ref{sec:softipod}, even with no outliers and $\bsbX=\bsbI$, a constant $\lambda$ independent of $n$ is
far from optimal~\cite{Donoho}.
It is also hard, in robust regressions, to select the cutoff value $\eta$ at which to identify outliers, because the residual distribution is usually unknown.
\citeasnoun{GYeff} base an asymptotically efficient choice for $\eta$ on a Kolmogorov-Smirnov statistic, but they need to assume that the standardized robust residuals are IID $\dnorm(0,1)$.

For $\Theta$-IPOD, 
these two issues correspond to the problem of tuning $\lambda$. 
In most $M$-estimators of practical interest, the $\psi$ function satisfies $\psi(t/\sigma;\lambda) \propto \psi(t;\lambda\sigma)$. Then we may set $\sigma=1$ and carefully tune $\lambda$.  Even with this simplification, a \emph{data-dependent} $\lambda$ is difficult to choose using cross-validation on the $(\bsbX, \bsby)$ data. The reason is that a large prediction error can indicate either a suboptimal $\bsbb$ or an outlier. Fortunately, turning to the reduced model \eqref{redmodel} greatly mitigates the problem because all (transformed) observations are \emph{clean}. Then BIC can be applied to $(\bsbA, \tilde\bsby)$ if the proportion of outliers is not large.

Specifically, because all candidate estimates lie along the $\Theta$-IPOD solution path, we design a {local} BIC to apply BIC on a proper local interval of the degrees of freedom (DF).
First we generate the hard-IPOD solution path by decreasing $\lambda$ from $\|(\bsbI - \bsbH)\bsby./\sqrt{\mbox{diag}(\bsbI-\bsbH)}\|_\infty$ to $0$.
Given $\lambda$ and the corresponding estimate $\hat \bsbg(\lambda)$, let $nz(\lambda) =\{ i: \hat \bsbg_i(\lambda)\neq 0\}$. We rely on model \eqref{redmodel} and study its variable selection  to give the correct form of BIC.
For hard-IPOD, $\hat \bsbg_{nz}$ is an OLS estimate with one parameter per detected outlier and the degrees of freedom are given by $\mbox{DF}(\lambda)=|nz(\lambda)|$.
We use BIC with a slight modification:
\begin{align}\label{eq:bicstar}
\mbox{BIC}^*(\lambda) = m \log(\mbox{RSS}/m) + k(\log(m)+1),
\end{align}
where  $m=n-p$, $\mbox{RSS}=\|\tilde\bsby - \bsbA\hat\bsbg\|_2^2 = \| (\bsbI-\bsbH)(\bsby - \hat\bsbg)\|_2^2$ and $k(\lambda)=\mbox{DF}(\lambda)+1$. We have taken $k=\mbox{DF}+1$ to account for the noise scale parameter, and BIC$^*$ uses $\log(m)+1$ instead of BIC's $\log(m)$ because we have found this change to be better empirically. Note that the sample size $m$ here is smaller than the dimension of $\bsbg$. 
According to \citeasnoun{chenchen}, similar modifications are necessary to preserve the model selection properties of BIC for problems with fewer observations than predictors.
We would like to apply  BIC$^*$ on a proper local interval of DF given by $[\nu_L, \nu_U]$.
We take $\nu_U\leq n/2$, assuming that the proportion of outliers is under $50\%$.  The $(\mbox{DF}(\lambda),\mbox{BIC}^*(\lambda))$ curve sometimes has narrow local minima near the ends of the $\lambda$ range. To counter that effect
we fit a smoothing spline to the set of data points $(\mbox{DF}(\lambda), \mbox{BIC}^*(\lambda))$ and chose the local minimum with the {largest} neighborhood. The neighborhood size can be determined using the local maxima of the smoothing spline. Of course there may be other reasonable ways to counter that problem.

We carried out simulation experiments to test the performance of the {tuned} $\Theta$-IPOD.
The matrix $\bsbX$ was generated
the same way as in Section~\ref{sec:compu}
using  dimension  $p\in\{15, 50\}$,  $n=1000$ observations of which the first
$O\in\{200, 150, 100, 50, 10\}$ were outliers at the highly leveraged location
given by $L\in\{15,20\}$ times a vector of $1$s.
The outliers were generated by a mean shift
$\bsbg=(\{5\}^{O}, \{0\}^{n-O})^\tran$ added to $\bsby$.
Because $\Theta$-IPOD is
affine, regression, and scale equivariant and the outliers are
at a single $x$ value,
we may set $\bsbb=\bsb{0}$ without loss of generality.
The intercept term is always included in the modeling.

Five outlier detection methods were considered for comparison: hard-IPOD (tuned), MM-estimator, 
~\possessivecite{GYeff}  fully efficient one-step procedure (denoted by GY), 
the compound estimator  S1S, and the LTS. 
(The direct procedures, such as~\citeasnoun{HS97},~\citeasnoun{PY95}, behaved poorly and their results were not reported.) The S-PLUS {\sc{{\tt Robust}}} library provides implementations of MM, GY, and LTS; for the implementation of S1S, we refer to~\citeasnoun{Wilbook}. {\sc{{\tt Robust}}} also provides a default initial estimate with  high breakdown point (not necessarily efficient), which is used in the first three algorithms in our experiments: when $p=15$, it is the S-estimate computed via random resampling; when $p=50$, it is the estimate from the fast PY procedure. 
Since it is well known that the initial estimate is outperformed by MM in terms of estimation efficiency and robustness, both theoretically and empirically~\cite{mmest,fastS,PYfast}, its results are not reported. (All of these routines are available for the {\tt R} language~\cite{Rlang} as well. The package {\tt robust} provides an {\tt R} version of the Insightful {\tt Robust} library.)

All methods apart from $\Theta$-IPOD require a cutoff value to identify which residuals are outliers.
We applied the fully efficient procedure which performs at least as well as the fixed choice of $\eta=2.5$ in various situations~\cite{GYeff}.



Each model was simulated $100$ times. We report outlier identification results for each algorithm using three 
benchmark measures:\par\noindent
\begin{tabular}{ll}
{\bf M} & the mean masking probability (fraction of undetected true outliers),\\
{\bf S} & the mean swamping probability (fraction of good points labeled as outliers),\\
{\bf JD}& the joint outlier detection rate (fraction of simulations with $0$ masking).
\end{tabular}\\

In outlier detection, masking is more serious than swamping. The former can cause gross distortions while
the latter is often just a matter of lost efficiency.

Ideally, $\mbox{M}\approx 0$, $\mbox{S}\approx 0$, and
$\mbox{JD}\approx 100\%$.
JD is the most important measure on easier problems while M makes the most sense for hard problems.
The simulation results are summarized in Tables~\ref{simutable1} and~\ref{simutable2}. Figures~\ref{figsimu_mjds_p15} and~\ref{figsimu_mjds_p50} present M and JD for $p=15$ and $50$ respectively. While our main purpose is to identify outliers, a robust coefficient estimate $\hat\bsbb$ can be easily obtained from $\Theta$-IPOD.
The MSE in $\bsbb$ for $p=15$ is shown in Figure \ref{figsimu_coefferrs}.
All methods had small slope errors, though $\Theta$-IPOD
and GY performed best. The results for $p=50$ (not shown) are similar.

\begin{figure}
\centering
\includegraphics[width=\hsize]{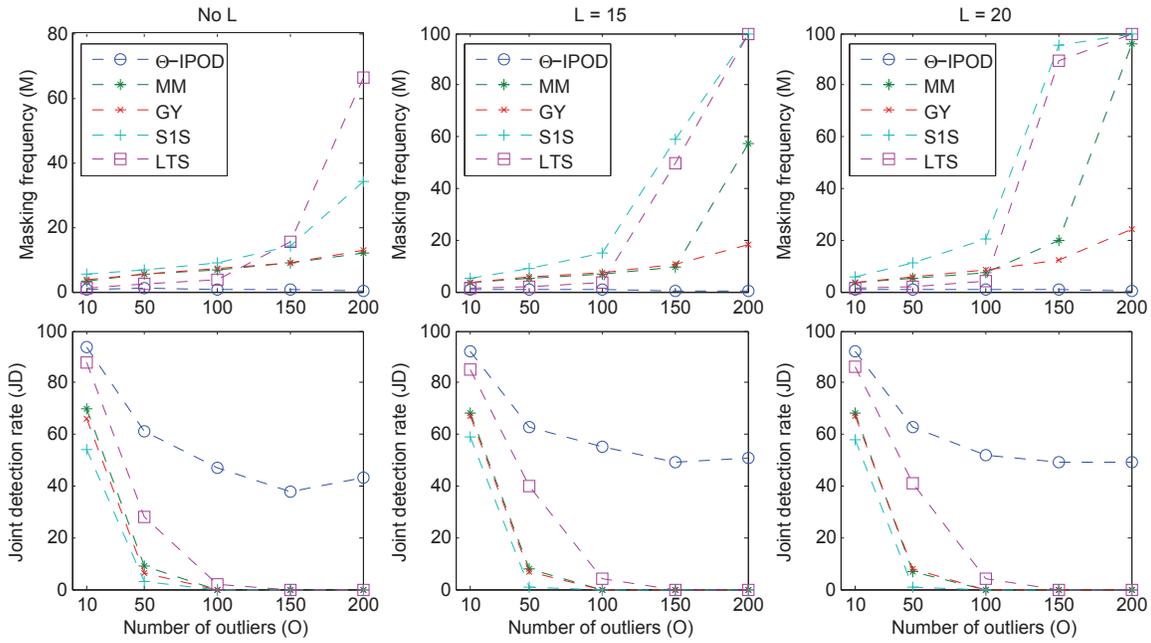}
\caption[]{\small{Masking (M) and joint detection (JD) results when $p=15$.}}
\label{figsimu_mjds_p15}
\end{figure}
\begin{figure}
\centering
\includegraphics[width=\hsize]{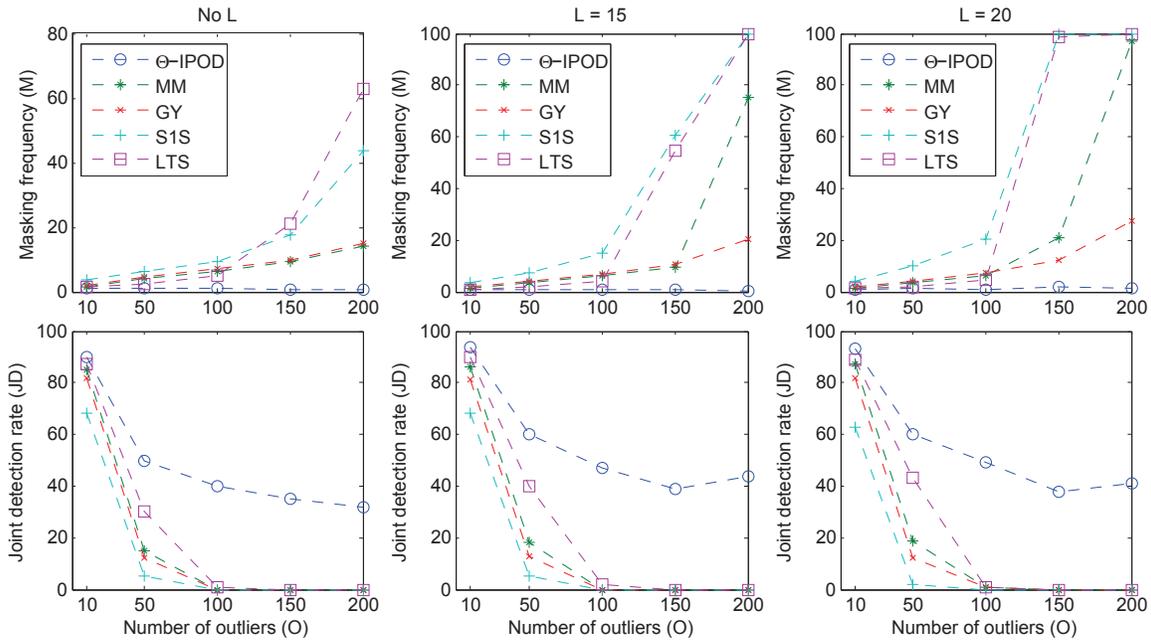}
\caption[]{\small{Masking (M) and joint detection (JD) when $p=50$.}}
\label{figsimu_mjds_p50}
\end{figure}

\begin{table}[ht]
\newcommand{\phm}{\phantom{-}}
\centering
\small{
\begin{tabular}{llrrrrrrrrrrrrrrr}
\hline

&&
\multicolumn{3}{c}{O = 200}&
\multicolumn{3}{c}{O = 100}&
\multicolumn{3}{c}{O = 50}&
\multicolumn{3}{c}{O = 20}&
\multicolumn{3}{c}{O = 10}\\
\hline
&
&{ JD}&{M\phm}&{ S}
&{ JD}&{M\phm}&{ S}
&{ JD}&{M\phm}&{ S}
&{ JD}&{M\phm}&{ S}
&{ JD}&{M\phm}&{ S}\\
\hline
\multicolumn{16}{l}{No $L$:}\\
&{IPOD}&  43 & 0.4 & 2.1 & 38 & 0.6 & 1.6  & 47 & 0.8 &  1.2  & 61 & 0.9 & 0.9  & 94 & 0.6 & 0.7 \\
&{MM}  &  0 & 11.9 & 0.0 & 0 & 8.8 & 0.0 & 0 & 6.9 & 0.1 & 9 & 5.2 & 0.1 & 70 & 3.4 & 0.3 \\
&{GY}  &  0 & 12.7 & 0.0 & 0 & 9.1 & 0.0 & 0 & 7.2 & 0.1 & 6 & 5.5 & 0.1 & 66 & 3.9 & 0.3 \\
&{S1S} &  0 & 34.3 & 0.0 & 0 & 14.3 & 0.1 & 0 & 8.9 & 0.1 & 3 & 6.6 & 0.1 & 54 & 5.5 & 0.2 \\
&{LTS} &  0 & 66.3 & 0.0 & 0 & 15.4 & 0.1 & 2 & 3.7 & 0.3 & 28 & 2.3 & 0.5 & 88 & 1.2 & 0.7 \\
\hline
\multicolumn{16}{l}{L=15:}\\
&{IPOD}&  51 & 0.4 & 2.2 & 49 & 0.5 & 1.6 & 55 & 0.6 & 1.2 & 63 & 0.8 & 0.9 & 92 & 0.8 & 0.7 \\
&{MM}  &  0 & 57.1 & 1.2 & 0 & 9.7 & 0.1 & 0 & 7.0 & 0.1 & 8 & 5.1 & 0.1 & 68 & 3.3 & 0.3 \\
&{GY}  &  0 & 18.5 & 0.0 & 0 & 10.6 & 0.1 & 0 & 7.6 & 0.1 & 7 & 5.4 & 0.1 & 67 & 3.6 & 0.3 \\
&{S1S} &  0 & 100.0 & 0.0 & 0 & 58.9 & 0.3 & 0 & 15.1 & 0.2 &1  & 8.9 & 0.2 & 59 & 5.1 & 0.2 \\
&{LTS} &  0 & 99.9 & 0.1 & 0 & 49.6 & 0.1 & 4 & 3.4 & 0.3 & 40 & 1.8 & 0.5 & 85 & 1.5 & 0.6 \\
\hline
\multicolumn{16}{l}{L=20:}\\
&{IPOD}&  49 & 0.4 & 2.1 & 49 & 0.6 & 1.6 & 52 & 0.7 & 1.2 & 63 & 0.9 & 0.9 & 92 & 0.8 & 0.7 \\
&{MM}  &  0 & 96.0 & 1.3 & 0 & 20.0 & 0.2 & 0 & 7.3 & 0.1 & 7 & 5.3 & 0.1 & 68 & 3.4 & 0.3 \\
&{GY}  &  0 & 24.1 & 0.1 & 0 & 12.2 & 0.1 & 0 & 8.2 & 0.1 &8  & 5.6 & 0.1 & 67 & 3.6 & 0.3 \\
&{S1S} &  0 & 100.0 & 0.1 & 0 & 95.5 & 0.1 & 0 & 20.1 & 0.2 & 1 & 11.0 & 0.2 & 58 & 5.7 & 0.2 \\
&{LTS} &  0 & 100.0 & 0.5 & 0 & 89.5 & 0.1 & 4 & 4.0 & 0.3 & 41 & 2.1 & 0.5 & 86 & 1.4 & 0.7 \\
\hline
\end{tabular}
}
\caption{\small{Outlier identification results on simulated data with $p=15$.
Five methods are compared: hard-IPOD (tuned), MM-estimator,  Gervini \& Yohai's fully efficient one-step procedure, S1S, and LTS.
Leverage $L$ and number of outliers $O$ are described in the text
as are the quality measures JD (joint detection),
M (masking) and S (swamping).
}}
\label{simutable1}
\end{table}

\begin{table}[ht]
\newcommand{\phm}{\phantom{-}}
\centering
\small{
\begin{tabular}{llrrrrrrrrrrrrrrr}
\hline
&&
\multicolumn{3}{c}{O = 200}&
\multicolumn{3}{c}{O = 100}&
\multicolumn{3}{c}{O = 50}&
\multicolumn{3}{c}{O = 20}&
\multicolumn{3}{c}{O = 10}\\
\hline
&
&{ JD}&{M\phm}&{ S}
&{ JD}&{M\phm}&{ S}
&{ JD}&{M\phm}&{ S}
&{ JD}&{M\phm}&{ S}
&{ JD}&{M\phm}&{ S}\\
\hline
\multicolumn{16}{l}{No $L$:}\\
&{IPOD}&  32 & 0.6 & 2.4 &  35 & 0.7 & 1.7 &  40 & 1.0 & 1.3 &  50 & 1.3 & 0.9 &  90 & 1 & 0.7 \\
&{MM}  &   0 & 13.9 & 0.0 &  0 & 9.2 & 0.1 &  0 & 6.4 & 0.1 &  15 & 4.0 & 0.3 &  85 & 1.5 & 0.6 \\
&{GY}  &  0 & 15.0 & 0.0 &  0 & 10.0 & 0.1 &  0 & 7.0 & 0.1 &  12 & 4.4 & 0.2 &  82 & 2 & 0.5 \\
&{S1S} &  0 & 43.8 & 0.1 &  0 & 17.8 & 0.1 &  0 & 9.3 & 0.1 &  5 & 6.1 & 0.2 &  68 & 3.7 & 0.4 \\
&{LTS} &  0 & 62.8 & 0.0 &  0 & 20.9 & 0.1 &  1 & 4.9 & 0.3 &  20 & 2.3 & 0.7 &  87 & 1.4 & 1.1 \\
\hline
\multicolumn{16}{l}{L=15:}\\
&{IPOD}& 44 & 0.5 & 2.4 &  39 & 0.7 & 1.7 &  47 & 0.9 & 1.3 &  60 & 1.1 & 0.9 &  94 & 0.6 & 0.7 \\
&{MM}  &  0 & 75.1 & 2.0 &  0 & 9.6 & 0.1 &  0 & 6.2 & 0.1 &  18 & 3.6 & 0.3 &  86 & 1.4 & 0.6 \\
&{GY}  &  0 & 20.3 & 0.1 &  0 & 10.5 & 0.1 &  0 & 6.7 & 0.1 &  13 & 3.9 & 0.2 &  81 & 1.9 & 0.5 \\
&{S1S} &  0 & 100.0 & 0.0 &  0 & 60.8 & 0.3 &  0 & 15.1 & 0.2 &  5 & 7.5 & 0.2 &  68 & 3.7 & 0.4 \\
&{LTS} &  0 & 100.0 & 0.7 &  0 & 54.4 & 0.1 &  2 & 4.0 & 0.3 &  40 & 2.0 & 0.7 &  90 & 1 & 1.0 \\
\hline
\multicolumn{16}{l}{L=20:}\\
&{IPOD}&  41 & 1.5 & 2.4 &  38 & 1.8 & 1.7 &  49 & 0.9 & 1.3 &  60 & 1.2 & 0.9 &  93 & 0.7 & 0.7 \\
&{MM}  &  0 & 97.1 & 1.3 &  0 & 21.2 & 0.3 &  1 & 6.4 & 0.1 &  19 & 3.7 & 0.3 &  87 & 1.3 & 0.6 \\
&{GY}  &  0 & 27.4 & 0.1 &  0 & 12.4 & 0.1 &  1 & 7.2 & 0.1 &  12 & 4.1 & 0.2 &  82 & 1.9 & 0.5 \\
&{S1S} &  0 & 100.0 & 0.3 &  0 & 99.6 & 0.1 &  0 & 20.2 & 0.2 &  2 & 10.1 & 0.3 &  63 & 4.1 & 0.4 \\
&{LTS} &  0 & 100.0 & 1.4 &  0 & 98.8 & 0.4 &  1 & 4.6 & 0.3 &  43 & 1.8 & 0.8 &  89 & 1.1 & 1.1 \\
\hline
\end{tabular}
}
\caption{\small{Outlier identification results on the simulation data where $p=50$.
Methods, conditions and results are the same as in Table~\ref{simutable1}.}}
\label{simutable2}
\end{table}

MM and GY are  two standard methods provided by the S-PLUS {\tt Robust} library. Nevertheless, as seen from the Tables \ref{simutable1} and \ref{simutable2}, the MM-estimator, though perhaps most popular in robust analysis, does not yield good identification results when the outliers have high leverage values and the number of outliers is not small (for example, $O=200$, $L=15, 20$). GY improves MM a lot in this situation and gives similar results otherwise.
The experiments also show that S1S behaves poorly in outlier detection and LTS works better in the presence of $\leq 5\%$ outliers. Unfortunately, all four methods have high masking probabilities and very low joint identification rates, which become worse for large $p$. The $\Theta$-IPOD method dominates them significantly for masking and joint detection.

To judge statistical significance of these MC results
we constructed paired $t$ statistics based on our
$100$ replicates. The numerator in each was the
number of outliers missed by a competing method
minus the number missed by $\Theta$-IPOD; the
denominator was the standard error of the numerator.
Most of the $t$-statistics were larger than $3$
and grew rapidly with $O$. LTS versus $\Theta$-IPOD
had the smallest $t$-statistic ($1.2$ for $O=15$)
but also the largest ($\sim 1700$ for $O=200$).
While $\Theta$-IPOD does much better on masking,
it is slightly worse for swamping. This is an acceptable
tradeoff because masking causes far more harm.




\begin{figure}
\centering
\includegraphics[width=\hsize]{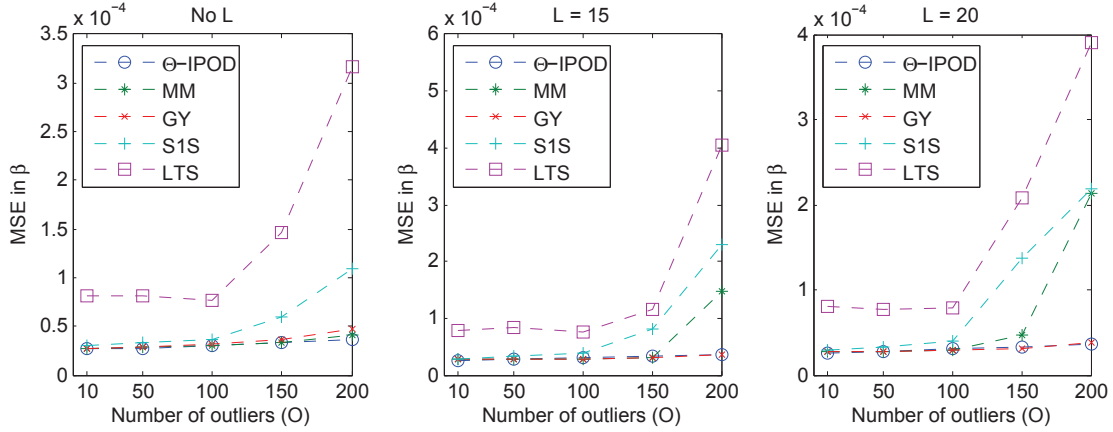}
\caption[]{\small{Coefficient estimation errors when $p=15$.}}
\label{figsimu_coefferrs}
\end{figure}

\section{Outlier Detection with $p>n$}\label{sec:largep}
Here we extend outlier detection to problems with $p>n$
including of course some {high-dimensional} problems.
This context is more challenging but has diverse modern applications in signal processing, bioinformatics, finance, and neuroscience among others. Performing the task of outlier identification for data with $p>n$ or even $p\gg n$ goes beyond the traditional robust analysis which requires a large number of observations relative  to the dimensionality.

The large-$p$ version of $\Theta$-IPOD is indeed possible under our general framework  \eqref{msomodel}. If $\bsbb\in \real^p$ is also assumed to be \emph{sparse} in the mean shift outlier model, one can directly work in an augmented data space $\bsby\sim \bsb{B} \bsb{\xi}$ with  $\bsb{B}= [\bsbX \ \bsbI]$ 
and $\bsb{\xi}=\left[\begin{array}{c} \bsbb \\\bsbg\end{array}\right]$ to study this large-$p$ sparsity problem.
Concretely, the TISP technique 
can be used to derive the  $\Theta$-IPOD iteration $\bsbxi^{(j+1)} = \Theta(\bsb{B}^\tran  \bsby/k_0^2  + (\bsbI-\bsb{B}^\tran\bsb{B}/k_0^2)\bsbxi^{(j)}; {\lambda}/k_0^2)$, or
\begin{equation}\label{IPOD-largep}
\begin{split}
\bsbb^{(j+1)} &\gets \Theta\left(\bsbb^{(j)} + \frac{1}{k_0^2}\bsbX^\tran \bsby  - \frac{1}{k_0^2}\bsbX^\tran (\bsbX\bsbb^{(j)}+\bsbg^{(j)}); \frac{\lambda}{k_0^2}\right)\quad\text{and}\\
\bsbg^{(j+1)} &\gets \Theta\left(\left(1-\frac{1}{k_0^2}\right)\bsbg^{(j)}+ \frac{1}{k_0^2}\bsby - \frac{1}{k_0^2}\bsbX\bsbb^{(j)}; \frac{\lambda}{k_0^2}\right).
\end{split}
\end{equation}
The convergence of~\eqref{IPOD-largep} is guaranteed  even for  $n<p$~\cite{SheTISP}, as long as $[\bsbX \ \bsbI]$ is scaled properly,
which here amounts to taking $k_0\geq \sqrt{\|\bsbX\|_2^2+1}$. Let  $\hat\bsbb$ and $\hat\bsbg$ denote the estimates from \eqref{IPOD-largep}. The nonzero components of $\hat\bsbb$  locate the relevant predictors while $\hat\bsbg$ identifies the outliers and measures their outlyingness. (We could also apply \eqref{IPOD-largep} when $p<n$ for simultaneous variable selection and outlier detection.)
Not surprisingly, $L_1$ (or soft-thresholding)  fails again for this challenging sparsity problem and thresholdings corresponding to redescending $\psi$'s should be used in \eqref{IPOD-largep}.
It is natural to consider the elastic net~\cite{ZouHas} for this problem as well. But the elastic net has a convex criterion, and it can be shown to have breakdown zero.

Since it is usually unknown how severe the outliers are, one may adopt the following hybrid of hard-thresholding and ridge-thresholding
\begin{eqnarray}
\Theta(x;\lambda,\eta)=\begin{cases} 0, & \mbox{if } |x|< \lambda\\
\frac{x}{1+\eta}, & \mbox{if }  |x|\geq\lambda \end{cases} \label{tisp:hybridthfunc}
\end{eqnarray}
referred to as the hybrid-thresholding in \citeasnoun{SheTISP}, or hard-ridge thresholding in this paper.
The corresponding penalty from the three-step construction is
$$P_\Theta(x;\lambda, \eta)=\begin{cases} -\frac{1}{2} x^2 + \lambda |x|, &\mbox{if } |x| < \frac{\lambda}{1+\eta}\\ \frac{1}{2} \eta x^2 +\frac{1}{2}\frac{\lambda^2}{1+\eta}, &\mbox{if } |x| \geq \frac{\lambda}{1+\eta}. \end{cases}
$$
Setting
$q(x;\lambda, \eta)=\frac{1+\eta}{2} (|x|-\lambda)^2  1_{0 < |x| < \lambda}$,
we get an alternative penalty 
\begin{align}
P(x; \lambda, \eta) = \frac{1}{2}\frac{\lambda^2}{1+\eta} 1_{x\neq0} + \frac12\eta x^2. \label{hybridpen2}
\end{align}
From \eqref{hybridpen2}, we see that
hard-ridge thresholding successfully fuses the $L_0$-penalty 
and the $L_2$-penalty (ridge-penalty) via $\Theta$-thresholding.
The $L_0$ portion induces sparsity, while the $L_2$ portion
shrinks $\bsbb$.
The difference from hard thresholding is as follows. When hard thresholding makes $\hat\gamma_i\ne 0$ we get $\hat\gamma_i=r_i$, removing any influence of observation $i$ on $\hat\beta$. With hard-ridge thresholding, the influence of the observation can be removed partially, with the extent of removal controlled by  $\eta$. This is helpful for nonzero but  small $|\gamma_i|$ corresponding to mild outlyingness. In addition, the $L_2$ shrinkage also plays a role in estimation and prediction.

The large $p$ setting brings a computational challenge.
Because the large-$p$ sparse $\Theta$-IPOD 
can be very slow, we used the following proportional $\Theta$-IPOD to screen out some `nuisance dimensions' with coefficients being exactly zero (due to our sparsity assumption).
Concretely, at each update in~\eqref{IPOD-largep},
$\lambda$ was chosen to get precisely $\alpha n$
nonzero components in the new $(\bsbb^{(j+1)}, \bsbg^{(j+1)})$.
We used $\alpha=0.75$ though other choices could be made.
Proportional $\Theta$-IPOD yields at most $\alpha n$ candidates predictors to have nonzero coefficients, making the problem simpler. Next, we  run \eqref{IPOD-largep} with just those predictors, getting a solution path in $\lambda$ and choosing $\lambda$ by BIC$^*$.
In proportional $\Theta$-IPOD,
a variable that gets killed 
at one iteration may reappear in the fit at a later iteration. 
This is quite different from independent screenings based on marginal statistics like FDR~\cite{fdr} or SIS~\cite{fanlv}. 

For our example we used the proportional hard-ridge IPOD. 
Specifically, for $\eta$ in a small grid
of values, we ran proportional $\Theta$-IPOD
using the hard-ridge thresholding function $\Theta$.
We chose $\eta$ by BIC$^*$.


We perform the joint robust variable selection and outlier detection on the sugar data of~\citeasnoun**{sugarpaper}. The data set concerns NIR spectroscopy of compositions of three sugars in aqueous solution. We concentrate on glucose (sugar 2) as the response variable.
The predictors are second derivative spectra of 700 absorbances at frequencies corresponding to wavelengths of 1100nm to 2498nm in steps of 2nm. There are 125 samples for model training.
A test set with 21 test samples is available. These 21 samples were specially designed to be difficult to predict, with compositions outside the range of the other 125. See~\citeasnoun{sugarbook} for details.

\citeasnoun{sugarpaper} used an 
MCMC Bayesian approach for variable selection, but only 160 equally spaced wavelengths were analyzed due to the computational cost.
We used all $700$ wavelengths in fitting our sparse robust model that also estimates outliers. The problem size is $125\times (700+125)$.
After the screening via proportional hard-ridge-IPOD, we ran hard-ridge-IPOD and tuned it as follows.  First we set $\lambda =0$
which turns $\eta$ into an ordinary ridge parameter. We fit that ridge parameter $\eta$, obtaining $\eta^*$ by minimizing $\mbox{BIC}^*$ of Section \ref{sec:tuning}.
Then for each $\eta$ in the grid $\{0.5\eta^*,0.05\eta^*,0.005\eta^*\}$, we found $\lambda(\eta)$
to minimize $\mbox{BIC}^*$, and finally chose among the three $(\lambda(\eta), \eta)$ combinations to minimize $\mbox{BIC}^*$.
We think there is no reason to use $\eta>\eta^*$ and
a richer grid than the one we used would be reasonable,
but we chose a small grid for computational reasons.
Our experience is that even small $\eta$ values improve prediction.


\begin{figure}[h]
\centering
$\begin{array}{c@{\hspace{0.0in}}c}
\includegraphics[width=7cm]{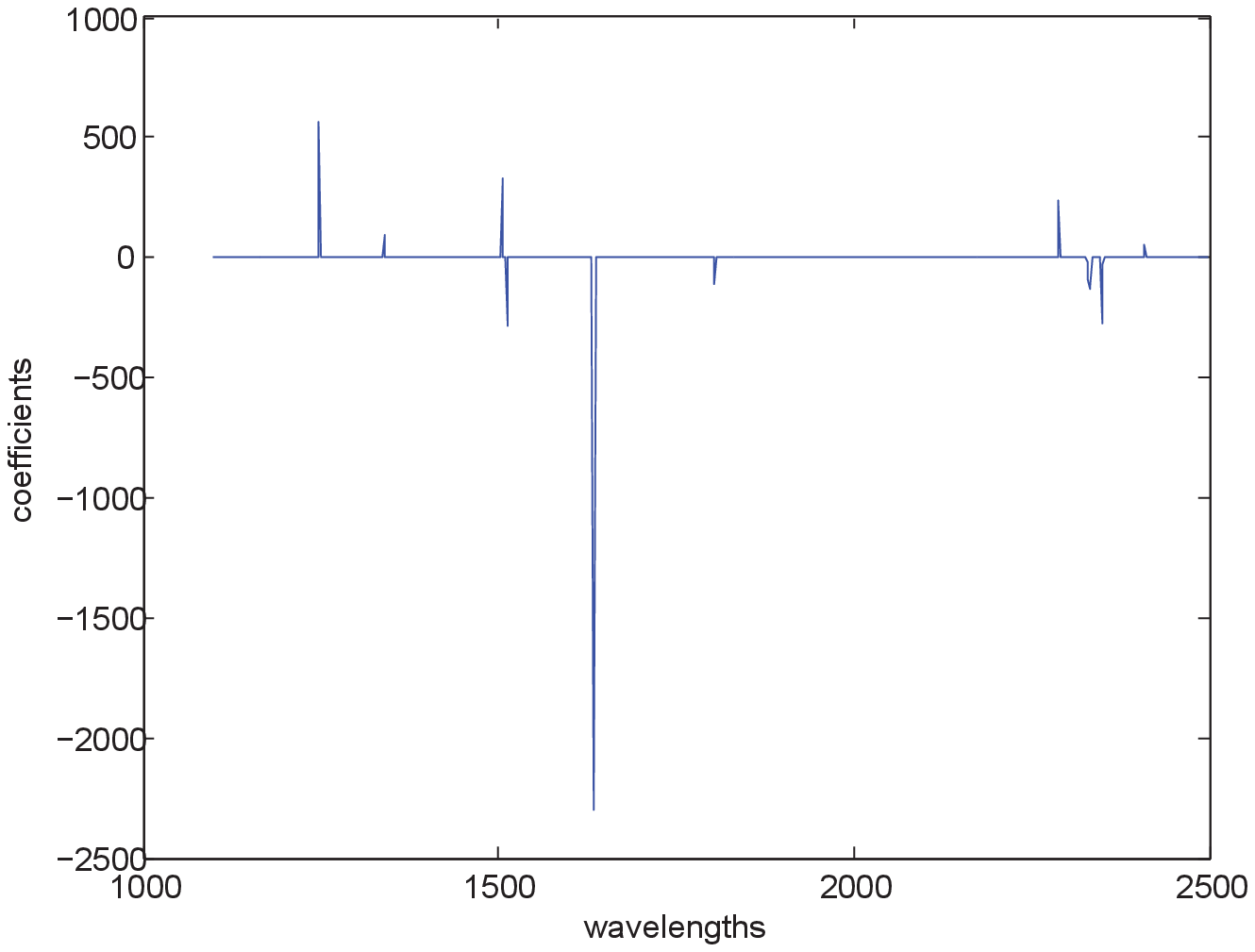} &
\includegraphics[width=6.4cm]{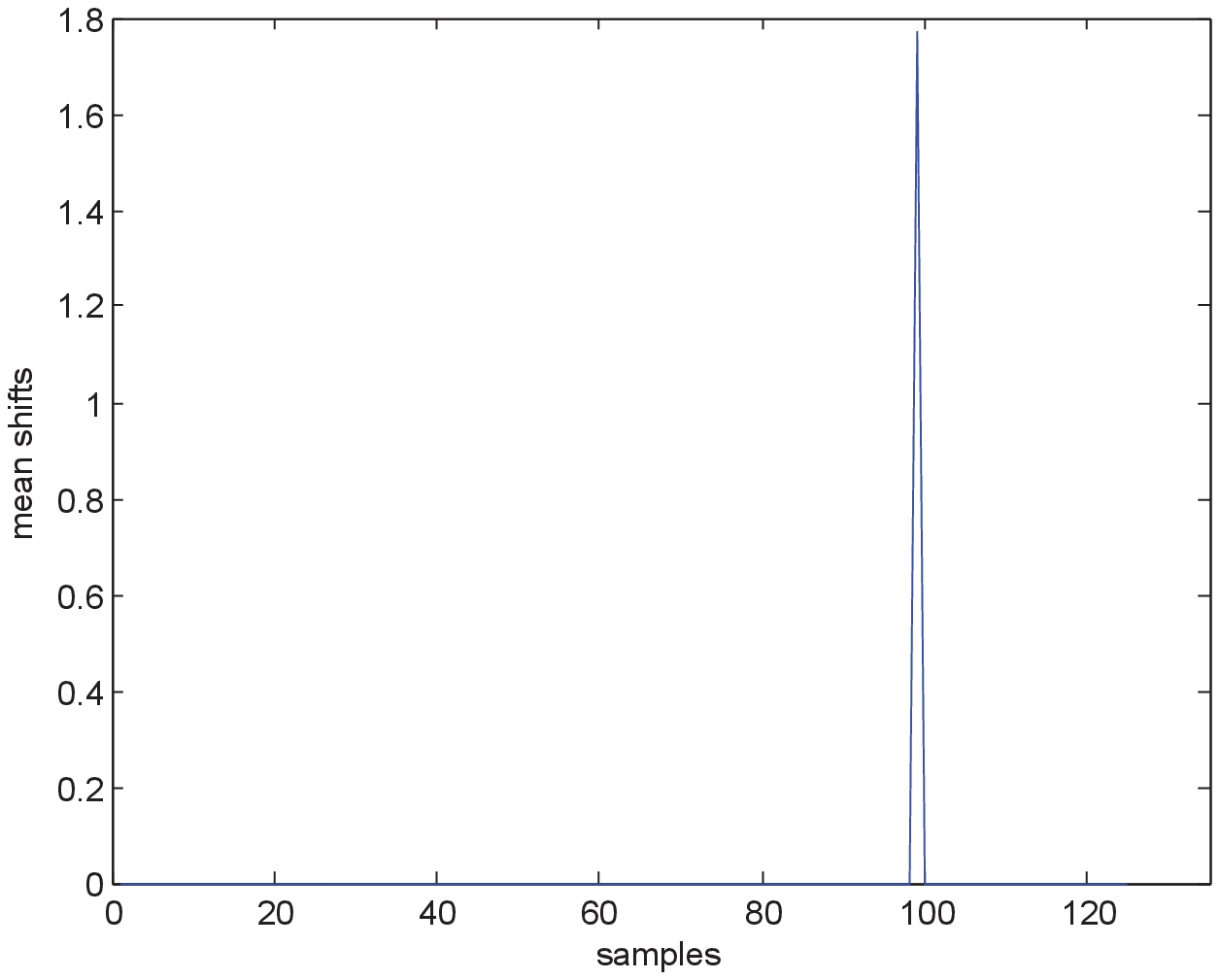}
\end{array}$
\caption[]{\small{Sparse coefficients and outlier shifts on the sugar data.}}\label{figcoeffshifts}
\end{figure}

Figure \ref{figcoeffshifts} plots the final estimates $\hat \bsbb$ and $\hat\bsbg$.
From the left panel, we see that the estimated model is sparse. It selects only $15$ 
 of the wavelengths.
The estimate $\hat\bsbg$  shown in
right side  of Figure~\ref{figcoeffshifts}
suggests that observation $99$ might be an outlier.
We found $\hat\gamma_{99}=1.7706$ 
 and
$r_{99}=y_{99}-\bsbx_{99}^\tran\hat \bsbb =1.7711$, 
which indicates that this observation is (almost) unused in the model fitting.
Our model has good prediction performance; the mean-squared error is 0.219 
on the test data, improving the reported MCMC results by about 39\%. 
The robust residual plot is shown in Figure \ref{figresids}.


\begin{figure}[t]
\centering
\includegraphics[width=9cm]{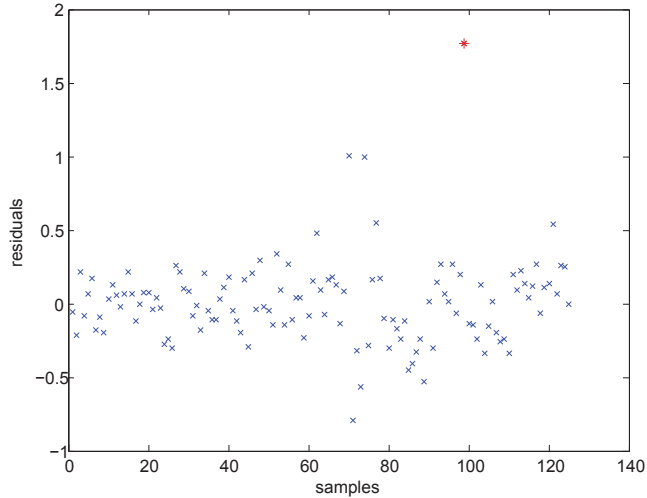}
\caption[]{\small{Residual plot from hard-ridge-IPOD on the sugar data. The apparent outlier,
observation $99$, is shown in red star.}}\label{figresids}
\end{figure}

\section{Discussion}\label{sec:discussion}
The main contribution of this paper is to
consider the class of $\Theta$-estimators~\cite{SheTISP}, under the mean shift outlier model assumption, defined by
the fixed point $\bsbg = \Theta(\bsbH \bsbg + (\bsbI-\bsbH)\bsby; \bsb{\lambda})$
which can be directly computed by $\Theta$-IPOD. With a good design of $\Theta$, we successfully identified the outliers as well as estimating the coefficients robustly.
This technique is associated with $M$-estimators, but gives a new characterization in form of penalized regressions. Furthermore, we successfully generalized this penalized/thresholding methodology to high-dimensional problems to accommodate and identify gross outliers in variable selection and coefficient estimation.

When outliers are also leverage points, the Gram matrix of the reduced model \eqref{redmodel}, i.e., $\bsbA^\tran\bsbA=\bsbI-\bsbH$, may demonstrate high correlation between clean observations and outliers for small $h_{i}$.
It is well known from recent advances of the lasso (e.g.\ the irrepresentable conditions~\cite{Zhao} and the sparse Riesz condition~\cite{ZhangHuang}) that the convex $L_1$-penalty encounters great trouble in this situation. Rather,  nonconvex penalties which correspond to redescending $\psi$'s must be applied, with a high breakdown point initial estimate obtained by, say, the fast-LTS~\cite{fastLTS} or fast-S~\cite{fastS} when $p$ is small, or  the fast PY procedure~\cite{PYfast} when $p$ is large.

In this framework, determining the  efficiency parameter in  $M$-estimators and choosing a cutoff value for outlier identification  are both accomplished by tuning the choice of $\bsb{\lambda}$. 
Our experience shows that adopting an appropriate data-dependent choice of $\bsb{\lambda}$ is crucial to guarantee good detection performance.

We close with a note of caution  on an important issue for the outlier detection literature.
We have found that our robust regression algorithms work better than competitors
on our simulated data sets and that they give the right answers on some small
well studied real data sets.  But our method and the others we study  all rely on a
preliminary robust fit. The most robust preliminary fits have a cost that grows
exponentially with dimension and for them $p=15$ is already large.
In high dimensional problems, the preliminary
fit of choice  is seemingly the PY procedure. 
It performed well
in our examples, but it does not necessarily have high breakdown.
Should the preliminary method fail, followup methods like $\Theta$-IPOD
or the others, may or may not correct it.  We have seen hard-IPOD work well from a non-robust start (e.g.\ the HBK problem) but it would  not be reasonable to expect this will always hold. 
We expect that $\Theta$-IPOD iterations will benefit from improvements
in preliminary robust fitting methods for high dimensional problems.
For low dimensional problems robust preliminary methods
like LTS are fast enough.

\appendix
\section{Proofs}
\noindent $\bullet$ \textbf{Proof of Proposition~\ref{propl1huber}}

\noindent
Since both Huber's method and the $L_1$-penalized regression are convex, it is sufficient to consider the KKT equations. In this proof, we define
$$
\Theta(t; \lambda) = \begin{cases} t-\lambda, &\mbox{if } t>\lambda \\ 0, &\mbox{if } |t|<\lambda \\ -t+\lambda, & \mbox{if } t<-\lambda  \end{cases}, \quad \psi(t; \lambda) = \begin{cases} \lambda, &\mbox{if } t>\lambda \\ t, &\mbox{if } |t|<\lambda \\ -\lambda, &\mbox{if } t<-\lambda  \end{cases}.
$$
Obviously, $\Theta(t; \lambda) + \psi(t; \lambda) = t.$
For simplicity, we use the same symbols for the vector versions of the two functions:
${\Theta}(\boldsymbol t; \lambda)=[\Theta(t_i, \lambda)]_{i=1}^n$, ${\psi}(\boldsymbol t; \lambda)=[\psi(t_i, \lambda)]_{i=1}^n$, $\forall \boldsymbol t\in \real^n$.

The KKT equations for Huber's estimate $(\hat \bsbb, \hat \sigma)$ are given by
\begin{eqnarray}
\bsbX^\tran   {\psi}\left(\frac{\bsby-\bsbX \bsbb}{ \sigma}; \lambda\right)&=&\bsb{0} \\
\frac{\partial}{\partial\sigma}
{{\left(\sum_{i=1}^n \rho\left( \frac{ y_i - \bsbX_i^\tran\bsbb}\sigma;\lambda \right)\sigma \right)}} + n c&=&0 \label{sigmaKKT}
\end{eqnarray}
Let
$
\hat \bsbr = \bsby-\bsbX\hat\bsbb, \hat G=\{i: |\hat r_i| < \lambda \hat \sigma\}, \mbox{ and } \hat O=\{i: |\hat r_i| > \lambda \hat \sigma\}. 
$
Since
$$
\frac{\partial}{\partial\sigma}
{{\left(2\rho(t/\sigma;\lambda)\sigma\right)}}=\begin{cases} -t^2/\sigma^2, &\mbox{if } |t|<\lambda\sigma\\ -\lambda^2, &\mbox{if } |t|>\lambda\sigma\end{cases}
$$
\eqref{sigmaKKT} becomes $nc = \lambda^2 |\hat O| + \sum_{i\in \hat G} \hat r_i^2/\hat \sigma^2$.
To summarize, $(\hat \bsbb, \hat \sigma)$ satisfies
\begin{eqnarray}
0&=&\bsbX^\tran  {\psi}\left(\frac{\bsby-\bsbX \hat\bsbb}{ \hat\sigma}; \lambda\right)\quad\text{and} \label{huberbeta}\\
\hat \sigma^2 &= & { \| \hat \bsbr_{\hat  G}\|_2^2}/ (cn - \lambda^2 |\hat O| ). \label{hubersigma}
\end{eqnarray}

Next, the joint KKT equations for $L_1$-penalized regression estimates $(\hat\bsbb, \hat\bsbg, \hat\sigma)$  are
\begin{align*}
\bsbX^\tran \bsbX\hat\bsbb &= \bsbX^\tran (\bsby-\hat\bsbg)\\
\hat\bsbg &= {\Theta}(\bsby-\bsbX\hat\bsbb;\lambda\hat\sigma)\quad\text{and}\\
0&=\|\bsby-\bsbX\hat\bsbb-\hat\bsbg\|_2^2 \cdot (-\frac{1}{\hat\sigma^2}) +nc,
\end{align*}
from which it follows that
$\bsbX^\tran  {\psi}\bigl(({\bsby-\bsbX \hat\bsbb})/{ \hat\sigma}; \lambda\bigr)=0$, and $\hat\sigma^2 = {\|\hat \bsbr-\hat\bsbg\|_2^2}/{nc}$.
But $\|\hat \bsbr-\hat\bsbg\|_2^2=\sum_{i\in\hat G} \hat r_i^2 + \lambda^2 \hat\sigma^2 |\hat O|$, so we obtain
\begin{align*}
\bsb{0}&=\bsbX^\tran  {\psi}\left(\frac{\bsby-\bsbX \hat\bsbb}{ \hat\sigma}; \lambda\right),
\quad\text{and}
\\
\hat \sigma^2 &=  { \| \hat \bsbr_{\hat  G}\|_2^2}/(cn - \lambda^2 |\hat O|), 
\end{align*}
which are exactly the same as \eqref{huberbeta} and \eqref{hubersigma}.
Therefore, \eqref{soft-ipodopt} leads to Huber's method with joint scale estimation. When $\sigma$ is fixed, it is not difficult to show that the two minimizations still yield the same $\bsbb$-estimate. 

Finally, Huber notices his method behaves poorly even for moderate leverage points and considers an improvement of using $c_1 \psi({\hat r_i}/{(c_2 \sigma)})$ to replace $\psi(\hat r_i/\sigma)$~\cite{Huberbook}. He claims, based on heuristic arguments, that $c_1=c_2=\sqrt{1-h_{i}}$ is a good choice. This is quite natural as seen from our new characterization. The regression matrix for $\bsbg$ in \eqref{soft-ipodopt} is actually $\bsbI-\bsbH$  with the column norms   given by $\sqrt{1-h_i}$. The weighted $L_1$-penalty of $\sum (\lambda \sqrt{1-h_i}) \times |\gamma_i|$ then leads to Huber's improvement, due to the fact that $K\psi(t/K;\lambda)=\psi(t; K\lambda)$ for $K > 0$. (However it cannot completely avoid masking and swamping unless a redescending $\psi$ is used.) \qed

\noindent $\bullet$ \textbf{Proof of Proposition~\ref{ipodm}}

\noindent
By definition, for any $\Theta$-IPOD estimate $(\hat \bsbb, \hat \bsbg)$, $\hat\bsbg$ is a fixed point of
$\bsbg = \Theta(\bsbH \bsbg + (\bsbI-\bsbH)\bsby; {\lambda})$, and $\hat\bsbb=(\bsbX^\tran\bsbX)^{-1} \bsbX^\tran (\bsby-\hat\bsbg)$. It follows that
\begin{align*}
\bsbX^\tran \psi(\bsby - \bsbX\hat\bsbb)& = \bsbX^\tran\psi(\bsby-\bsbH(\bsby-\hat\bsbg);\lambda)\\
&=\bsbX^\tran (((\bsbI-\bsbH)\bsby+\bsbH\hat\bsbg) - \Theta((\bsbI-\bsbH)\bsby+\bsbH\hat\bsbg); \lambda))\\
&= \bsbX^\tran ( (\bsbI-\bsbH)\bsby + \bsbH\hat\bsbg - \hat\bsbg) \\
&= \bsbX^\tran (\bsbI-\bsbH) \cdot (\bsby - \hat \bsbg) = \bsb{0},
\end{align*}
and so $\hat\bsbb$ is an $M$-estimate associated with $\psi$. \qed

\noindent $\bullet$ \textbf{Proof  of Theorem~\ref{thpen}}

\noindent
The second inequality in \eqref{ipodconvineq} is straightforward from the algorithm design. To show the first inequality is true, it is sufficient to prove the following lemma.
\begin{lemma}
\label{uniqsol-gen}
Given a thresholding rule $\Theta$, let $P$ be any function satisfying $P(\theta;\lambda)=P(0;\lambda)+P_{\Theta}(\theta; \lambda) + q(\theta; \lambda)$ where $q(\cdot; \lambda)$ is nonnegative and $q(\Theta(\theta;\lambda))=0$ for all $\theta$.  Then, the minimization problem $\min_{\theta}  (t-\theta)^2/2 + P(\theta;\lambda)$ has a unique optimal solution $\hat\theta=\Theta(t;\lambda)$ for every $t$ at which $\Theta(\cdot;\lambda)$ is continuous.
\end{lemma}
This is a generalization of Proposition 3.2 in~\citeasnoun{antrev}. Note that $P$ (and $P_{\Theta}$) may not be differentiable at 0 and may not be convex.

Without loss of generality, suppose $t> 0$. It suffices to consider $\theta\geq 0$ since $f(\theta)\geq f(-\theta), \forall \theta\geq 0$, where  $f(\theta)\equiv (t-\theta)^2/2 + P(\theta;\lambda)$. First, $\Theta^{-1}$ given by \eqref{defofthetainv} is well-defined. We have
\begin{eqnarray*}
f(\theta)-f(\Theta(t;\lambda))&=&\int_{\Theta(t;\lambda)}^{\theta} f'(u)\rd u + q(\theta; \lambda) - q(\Theta(t;\lambda)) \\
&=&\int_{\Theta(t;\lambda)}^{\theta} (u-t+\Theta^{-1}(u;\lambda)-u)\rd u+ q(\theta;\lambda)\\
&=&\int_{\Theta(t;\lambda)}^{\theta} (\Theta^{-1}(u;\lambda)-t)\rd u+ q(\theta; \lambda).
\end{eqnarray*}
Suppose $\theta>\Theta(t;\lambda)$. By definition $\Theta^{-1}(\theta;\lambda)\geq t$, and thus $f(\theta)\geq f(\Theta(t;\lambda))$. There must exist some $u\in[\Theta(t;\lambda), \theta)$ s.t. $\Theta^{-1}(u;\lambda)>t$. Otherwise we would have $\Theta(t';\lambda)>\theta\geq\Theta(t;\lambda)$ for any $t'>t$, and thus $\Theta(\cdot)$ would be discontinuous at $t$.
Noticing that $\Theta^{-1}(\cdot)$ is monotone, $\int_{\Theta(t;\lambda)}^{\theta} (\Theta^{-1}(u;\lambda)-t)\rd u>0$, or $f(\theta)>f(\Theta(t;\lambda))$. A similar reasoning applies to the case $\theta<\Theta(t;\lambda)$. The proof is now complete. \qed

\onehalfspacing
\bibliographystyle{ECA_jasa}
\bibliography{ipodbib}

\end{document}